\documentclass[12pt]{iopart}

\usepackage{hyperref} 
\usepackage{graphicx}
\usepackage{color}
\usepackage{dcolumn}
\usepackage{bm}
\usepackage{verbatim} 
\usepackage{cite} 
\usepackage[normalem]{ulem}

\begin{document}


    \long\def \cblu#1{\color{blue}#1}
    \long\def \cred#1{\color{red}#1}
    \long\def \cgre#1{\color{green}#1}
    \long\def \cpur#1{\color{purple}#1}

\newcommand{\eric}[1]{{\color{blue}#1}}
\newcommand{\guido}[1]{{\color{violet}#1}}
\newcommand{\matthias}[1]{{\color{blue}#1}}
\newcommand{\fabian}[1]{{\color{blue}#1}}
\newcommand{\di}[1]{{\color{blue}#1}}
\newcommand{\ericC}[1]{{\color{red}\textit{\textbf{Eric:} #1}}}
\newcommand{\guidoC}[1]{{\color{red}\textit{\textbf{Guido:} #1}}}
\newcommand{\matthiasC}[1]{{\color{red}\textit{\textbf{Matthias:} #1}}}
\newcommand{\fabianC}[1]{{\color{red}\textit{\textbf{Fabian:} #1}}}
\newcommand{\diC}[1]{{\color{red}\textit{\textbf{Di:} #1}}}

\def\FileRef{
\input FName
{
\newcount\hours
\newcount\minutes
\newcount\min
\hours=\time
\divide\hours by 60
\min=\hours
\multiply\min by 60
\minutes=\time
\
\advance\minutes by -\min
{\small\rm\em\the\month/\the\day/\the\year\ \the\hours:\the\minutes
\hskip0.125in{\tt\FName}
}
}}

\mathchardef\muchg="321D
\let\na=\nabla
\let\pa=\partial

\let\muchg=\gg

\let\t=\tilde
\let\ga=\alpha
\let\gb=\beta
\let\gc=\chi
\let\gd=\delta
\let\gD=\Delta
\let\ge=\epsilon
\let\gf=\varphi
\let\gg=\gamma
\let\gh=\eta
\let\gj=\phi
\let\gF=\Phi
\let\gk=\kappa
\let\gl=\lambda
\let\gL=\Lambda
\let\gm=\mu
\let\gn=\nu
\let\gp=\pi
\let\gq=\theta
\let\gr=\rho
\let\gs=\sigma
\let\gt=\tau
\let\gw=\omega
\let\gW=\Omega
\let\gx=\xi
\let\gy=\psi
\let\gY=\Psi
\let\gz=\zeta

\let\lbq=\label
\let\rfq=\ref
\let\na=\nabla
\def\daI{{\dot{I}}}
\def\dsq{{\dot{q}}}
\def\dgj{{\dot{\phi}}}

\def\bgs{\bar{\sigma}}
\def\bgh{\bar{\eta}}
\def\bgg{\bar{\gamma}}
\def\bgy{\bar{\psi}}
\def\bgF{\bar{\Phi}}
\def\bgY{\bar{\Psi}}

\def\baF{\bar{F}}
\def\bsj{\bar{j}}
\def\baJ{\bar{J}}
\def\bsp{\bar{p}}
\def\baP{\bar{P}}
\def\bsx{\bar{x}}

\def\hgj{\hat{\phi}}
\def\hgq{\hat{\theta}}

\def\HaT{\hat{T}}
\def\HaR{\hat{R}}
\def\Hsb{\hat{b}}
\def\Hsh{\hat{h}}
\def\Hsz{\hat{z}}

\let\gG=\Gamma
\def\taA{{\tilde{A}}}
\def\taB{{\tilde{B}}}
\def\taG{{\tilde{G}}}
\def\tsp{{\tilde{p}}}
\def\tsv{{\tilde{v}}}
\def\tgF{{\tilde{\Phi}}}

\def\wgx{{\bm{\xi}}}
\def\wgz{{\bm{\zeta}}}

\def\wse{{\bf e}}
\def\wsk{{\bf k}}
\def\wsi{{\bf i}}
\def\wsj{{\bf j}}
\def\wsl{{\bf l}}
\def\wsn{{\bf n}}
\def\wsp{{\bf p}}
\def\wsr{{\bf r}}
\def\wsu{{\bf u}}
\def\wsv{{\bf v}}
\def\wsx{{\bf x}}

\def\vaB{\vec{B}}
\def\vse{\vec{e}}
\def\vsh{\vec{h}}
\def\vsl{\vec{l}}
\def\vsv{\vec{v}}
\def\vgn{\vec{\nu}}
\def\vgk{\vec{\kappa}}
\def\vgt{\vec{\gt}}
\def\vgx{\vec{\xi}}
\def\vgz{\vec{\zeta}}

\def\waA{{\bf A}}
\def\waB{{\bf B}}
\def\waD{{\bf D}}
\def\waE{{\bf E}}
\def\waF{{\bf F}}
\def\waJ{{\bf J}}
\def\waV{{\bf V}}
\def\waX{{\bf X}}

\def\R#1#2{\frac{#1}{#2}}
\def\btbl{\begin{tabular}}
\def\etbl{\end{tabular}}
\def\bqbl{\begin{eqnarray}}
\def\eqbl{\end{eqnarray}}
\def\ebox#1{
  \begin{eqnarray}
    #1
\end{eqnarray}}


\def \cred#1{{\color{red}(#1)}}
\def \cblu#1{{\color{blue}#1}}

\title[Hot-tail electron impact on ablation]{Hot-tail electrons' impact on assimilation and injection penetration of D2 Shattered Pellet Injections}
\author{D. Hu$^{1}$, C. Liu$^{2}$ \& JOREK Team\footnote{See https://www.jorek.eu for a list of current team members}}
 \ead{hudi2@buaa.edu.cn}
\address{
$^1$Beihang University, No. 37 Xueyuan Road, Haidian District, 100191 Beijing, China.
}
\address{
$^2$Princeton Plasma Physics Laboratory, Princeton, NJ 08540, United States of America
}
\ead{hudi2@buaa.edu.cn}

\vspace{10pt}
\begin{indented}
\item[]\today
\end{indented}

\begin{abstract}
    The assimilation and the penetration of injected materials play significant roles in the mitigation efficiency of Shattered Pellet Injection (SPI) as a Disruption Mitigation System (DMS), hence an accurate description of the fragment ablation rate is desirable. Current mainstream 3D MHD codes modelling SPIs mostly assume instantaneous thermalization between the previously hot ambient electrons and the newly released cold electrons. This could result in underestimation of the ablation rate if the hot electron thermalization time is comparable or even longer than the fragment flying time across flux tubes, putting the validity of these 3D SPI simulations into question. To resolve this doubt, we hereby investigate the thermalization dynamics and the overall hot-electron impact. The finite-time collisional thermalization of hot-tail electrons in a rapidly cooling plasma, as well as the so-called ``self-limiting'' effect are considered. The former effect tends to deplete the colder population within a hot-tail species, while the latter is found to preferentially deplete the higher energy population. The combined result is found to cause an almost self-similar decay of the hot electron distribution function, while its shape does not deviate much from that of Maxwellian distribution and the mean energy does not change much during the thermalization process. Based on this observation, axisymmetric JOREK D2 SPI simulations were carried out with additional hot-tail contribution to evaluate their overall impact onto the injection assimilation and penetration. It is found that the hot-tail effect indeed causes enhanced assimilation and shallower penetration, although the overall effect depends on the exact injection configuration, with the slow injection showing negligible hot-tail effect while the fast single non-shattered pellet case shows drastic hot-tail ablation enhancement. For ITER-like SPI parameters, there is no significant deviation in the total assimilation, but some deviation in the injection penetration is observed for the fast injection velocity cases.
\end{abstract}

%
%
%
\maketitle
%
%

\section{Introduction}
\label{s:Intro}

Shattered Pellet Injection (SPI) is one of the main Disruption Mitigation Systems intended to mitigate strong Thermal Quench (TQ) heat flux as well as to suppress possible Current Quench (CQ) runaway electron current generation \cite{Lehnen2015JNM}. To achieve good mitigation efficiency, enough assimilation and penetration of the injected materials are essential \cite{Di2018NF,Di2021NF}, hence an accurate description of the fragment ablation is desirable. Usually, the ablation rate of the fragments could be conveniently estimated by considering a power-law of the electron temperature and density of the ambient plasma, as well as the fragment radius and perhaps the magnetic field \cite{Sergeev2006PPR,Parks2022XXX,Zhang2020NF,Bosviel2021POP}.

During a massive material injection scenario, however, the ambient plasma itself is rapidly cooling down as a result of the strong dilution and radiation, making the accurate description by the aforementioned power law difficult, especially considering the strong power dependence on the ambient electron temperature. Currently, main stream 3D MHD codes \cite{Di2018NF,Di2021NF,Kim2019POP} mostly consider the instantaneous thermalization between the target hot plasma and the cold plasma newly released from the ablation, so that the calculated ablation rate will decrease drastically shortly after the fragments arrive on a flux surface and cool it down. This, however, may not be accurate. For high electron temperature target plasma, the collisional thermalization time could be comparable or even longer than the flying time of the fragments, so that the ambient plasma ``seen'' by the fragments is never thermalized. Instead, a bulk cold electron species and a ``hot-tail'' electron species are present simultaneously. Since the hot electrons could better penetrate the shielding cloud around the fragments \cite{Pegourie2005PPCF}, the ablation should be dominated by the hot electron species. The current instantaneous thermalization treatment thus risks to underestimate the fragment ablation rate in a cooling plasma. Indeed, this finite thermalization time may even result in a persistent long-tail hot electron species \cite{Smith2008POP,Aleynikov2017NF}, and recent DIII-D experiments have shown evidence suggesting the creation of such hot-tail electrons after pellet injections and their enhancement of the pellet ablation \cite{Hollmann2021POP,Hollmann2021NF}.

To resolve this concern, we hereby investigate the finite thermalization time effect of the hot electrons after SPI, and compare the fragment assimilation and penetration considering the hot-tail electron effect against that from a instantaneously thermalized fluid consideration. To do this, we first look at the characteristics of hot electron distribution function during thermalization, then try to construct a fluid representation of these hot electrons from their distribution to be used in axisymmetric fluid simulations.

We consider two dominant effects during the thermalization, one is the collisional thermalization \cite{Smith2008POP} and the other is the so-called self-limiting effect \cite{Houlberg1988NF}. The former is found to preferentially deplete the colder population of the hot electron species, while the latter shows opposite preference. Using numerical methods described in Ref.\,\cite{Liu2018PRL} and Ref.\,\cite{Liu2018NF}, simulations of the hot electron distribution function find that, for hot enough initial temperature, the distribution function exhibits an almost self-similar depletion as they are assimilated into the cold, dense background electrons. The thermalization process thus shows quick density depletion of the hot electrons, while the mean energy of the hot electrons remains more or less unchanged. The shape of the hot electron distribution function is found to not deviate much from that of a Maxwellian during this process. This enable us to directly use existing power law \cite{Zhang2020NF} to calculate the hot-tail contribution to the ablation rate.

Utilizing above observations, axisymmetric JOREK \cite{Hoelzl2021NF} ITER SPI simulations with different injection configuration are carried out. The hot-tail electrons are considered by a simple 1D model where the hot electron density and temperature are functions of the normalized magnetic flux, and their evolutions are determined by the collisional thermalization and the self-limiting effect. The hot-tail electrons are found to indeed result in enhanced assimilation and shallower penetration of the fragments, however the strength of such effect varies depends on the exact SPI configuration. We find the injection velocity plays important role in the strength of the hot-tail effect, with a slow injection showing negligible hot-tail impact. Further, the compactness of the plume shows some self-regulating behavior. A more compact plume is found to increases the hot-tail effect while also increasing the self-limiting mitigation of it, thus the two partly cancel out each other. Using finer fragments to increase the fragment density on a given surface is found to increase the self-limiting mitigation of the hot-tail effect. Finally, a Unshattered Pellet case is also conducted to show that in the absence of the self-limiting effect and fragment dispersion, the hot-tail effect is tremendous and a simple instantaneously thermalized consideration will underestimate the assimilation while overestimate the injection penetration significantly in this extreme case.

The rest of the paper is arrange as the following. In Section \ref{s:DistributionFunction}, we investigate the hot electron thermalization process by looking at its distribution function evolution numerically, and fluid representation of hot electron density and mean energy evolution are constructed. Based on such representations, JOREK axisymmetric simulations are carried out in Section \ref{s:OverallImpact} to access the hot-tail electron contribution to the injection assimilation and penetration. Finally, conclusion and discussion are given in Section \ref{s:Conclusion}.

\section{Hot electron distribution function evolution during thermalization}
\label{s:DistributionFunction}

We present the numerical investigation of the hot-tail electron distribution function under deuterium SPI scenario. Specifically, in Section \ref{ss:CTnSL}, we introduce the collisional thermalization and the self-limiting process. In Section \ref{ss:NumericalThermalization}, simulation of the hot electron distribution function is carried out, and and the characteristic distribution evolution behavior for different initial temperature and characteristic fragment size is demonstrated. In Section \ref{ss:AnalyticalThreshold}, analytical modelling is carried out to obtain the hot electron density sink for coupling with fluid simulations in Section \ref{s:OverallImpact}.

\subsection{Collisional thermalization and the self-limiting effect}
\label{ss:CTnSL}

The distribution function evolution of the hot electron species within a bulk cold electron background, without consideration of the quasi-linear diffusion from wave-particle interaction and the avalanche mechanism, could be written as the following \cite{Liu2018PRL}:
\bqbl
\R{\pa f}{\pa t}
+\R{eE_\|}{m_ec}\left(\gx\R{\pa f}{\pa p}+\R{1-\gx^2}{p}\R{\pa f}{\pa \gx}\right)
+C[f]+\R{\pa}{\pa \wsp}\cdot\left(\waF_{rad}f\right)
=
0
.\eqbl
Here, $E_\|$ is the parallel electric field, $\gx\equiv p_\|/p$ is the cosine of the pitch angle, $m_e$ is the electron mass, $e$ is its charge, $C[\cdots]$ represents the Fokker-Planck test-particle collision operator \cite{Liu2011POP} and $\waF_{rad}$ is the synchrotron radiative drag \cite{Stahl2015PRL}. We have neglected the collision between the hot species, only focus on the collision between the hot and the cold electrons as well as the ions. With negligible parallel electric field, the hot electron distribution as a function of the normalized time $S$ could be described as \cite{Smith2008POP}:
\bqbl
\lbq{eq:SmithFunction}
f(v)
=
\R{4n_{e,hot}(0)}{\sqrt{\gp}v_{0}}\R{v^2}{v_{0}^2}\exp{\left[-\left(\R{v^3}{v_{0}^3}+S\right)^{2/3}\right]}
.\eqbl
Here, the initial thermal velocity $v_{0} = \sqrt{2T_{e,hot}(0)/m_e}=\sqrt{2\gq(0)}c$, the dimensionless parameter $\gq\equiv\R{T}{m_ec^2}$ is the normalized temperature. The normalized time is defined as $S\equiv 3s\R{c^3}{v_0^3}$, with $\R{ds}{dt}\equiv\R{e^4n_{e,cold}\ln{\gL}}{4\gp\ge_0^2m_e^2c^3}$. The number density of the hot and the cold electron species are $n_{e,hot}$ and $n_{e,cold}$ respectively. Integrating over the distribution function, one will find the time evolution of the hot electron number density could be analytically described as \cite{Aleynikov2017NF}:
\bqbl
\lbq{eq:SmithDensity}
\R{n_{e,hot}(S)}{n_{e,hot}(0)}
=
\R{2S^{1/3}}{\sqrt{\gp}}\exp{\left[-S^{2/3}\right]}+erfc\left(S^{1/3}\right)
.\eqbl
Here $erfc(\cdots)\equiv 1- erf(\cdots)$ is the complementary error function.
While the integration for the mean energy is more analytically challenging, numerical integration nonetheless shows the hot electron mean energy is rising over time instead of decreasing \cite{Aleynikov2017NF}, as would have been naively anticipated in a thermalization process. This is because the colder population in the hot electron species experiences higher collision frequency, thus is quickly assimilated into the bulk cold electron species, while the hotter population remains.
Hence the collisional thermalization process tends to create a hot-tail in the electron distribution which could potentially cause significant enhancement to the fragment ablation.

On the other hand, due to the high atom number density in solid fragments which is typically of $\mathcal{O}\left(10^{28}/m^3\right)$, the mean-free-path of the hot electrons with energy in the $10keV$ range is approximately of $\mathcal{O}\left(10^{-5}m\right)$ \cite{Houlberg1988NF,Pegourie2005PPCF}, much smaller than the characteristic size of the fragments \cite{Gebhart2020FST,Baylor2021NF}. For the small population of $\mathcal{O}\left(100keV\right)$ electrons, their mean-free-path is comparable with the size of the fragments. Even if they do not stop in the solid, significant amount of energy would be lost as they pass through. Hence we consider hot electrons colliding with the solid fragments lost and assimilated into the cold electron background.
Note that, according to the Neutral Gas Shielding (NGS) model \cite{Parks1977NF}, the majority of the parallel electron heat flux would be deposited within the neutral cloud surrounding the pellet, thus it would seem that using the cross-section of the neutral cloud instead of that of the solid fragment to consider the self-limiting effect is more appropriate. Since the neutral cloud radius is approximately 1.4 times that of the solid fragment \cite{Pegourie1993NF}, this would result in doubling of the self-limiting cross-section. However, since the electrons tangentially passing through the outer, less dense region of the neutral cloud would experience reduced line-integrated-density, it is not trivial to find an accurate description of the self-limiting cross-section considering the neutral gas contribution. Thus we hereby only consider the self-limiting effect from the solid fragment, which should be considered as a lower estimation.

For electrons with parallel velocity $v_\|$, each fragment sharing the same flux tube with these electrons then presents a sink in the distribution function with the form:
\bqbl
\left(\R{df}{dt}\right)_{sl}
=
-v\left|\gx\right|S_pf/\gD V
.\eqbl
Here, $S_p$ is the effective cross-section of the fragment and $\gD V$ is the volume of the flux tube. In principle, such a parallel velocity dependent sink would cause significant anisotropy in the momentum space. We however assume that the parallel electric field and the pitch angle scattering produced by magnetic perturbation or collision would tend to reduce such anisotropy. So that the self-limiting sink could be written as a function of velocity alone:
\bqbl
\lbq{eq:SLSink}
\left(\R{df}{dt}\right)_{sl}
=
-\R{v}{2}S_p f/\gD V
.\eqbl
Here the $1/2$ appears as a result of integration over the pitch angle.
This sink term will preferentially deplete the hotter population within the hot electron species, as opposed to the aforementioned collisional process. This is the so-called self-limiting effect \cite{Houlberg1988NF}. Note that if the electron distribution has a tail along the parallel direction like usually found in hot-tail runaway electron seed generation studies \cite{Petrov2021PPCF}, then Eq.\,(\rfq{eq:SLSink}) would underestimate the self-limiting sink.

Combining the collisional thermalization and the self-limiting effect, the hot electron distribution could be described as
\bqbl
\lbq{eq:HotEKinetic}
\R{\pa f}{\pa t}
+\R{eE_\|}{m_ec}\left(\gx\R{\pa f}{\pa p}+\R{1-\gx^2}{p}\R{\pa f}{\pa \gx}\right)
+C[f]+\R{\pa}{\pa \wsp}\cdot\left(\waF_{rad}f\right)
=
\sum_N{\left(\R{df}{dt}\right)_{sl}}
.\eqbl
Here, $\sum_N$ represents the summation over all fragments in the flux tube. Eq.\,(\rfq{eq:SLSink}) and Eq.\,(\rfq{eq:HotEKinetic}), accompanied by suitable choice of fragment cross-section and the flux tube volume, then form the governing equations for our following numerical investigation of the hot electron distribution function.

\subsection{Numerical simulation of hot electron thermalization}
\label{ss:NumericalThermalization}

We consider the slowing down of the hot electrons in a homogeneous bulk cold plasma background in the wake of the vanguard fragments after the deuterium SPI. Let the ions be cold and the background cold electrons temperature is constant at $T_{e,cold}=5eV$. The background cold electron density is set to be constant $n_{e,cold}=10^{21}/m^3$, in accordance with past numerical observations \cite{Di2018NF,Di2021NF}. As for the initial hot electron species, ITER-like parameter is considered with initial hot electron temperature $T_{e,hot}(0)$ ranging from $5keV$ to $20keV$, while $n_{e,hot}(0)=10^{20}/m^3$.
The local electric field is obtained from self-consistent iteration using the electron distribution function and the local current density, while the local current density evolution is described approximately by
\bqbl
\R{\pa J_\|}{\pa t}
=
\R{1}{\gm_0}\na_\bot^2E
\sim
-\R{E}{\gm_0 L_J^2}
.\eqbl
Here we have assumed that $\pa_t \gh$ is negligible, otherwise we have to also iterate over time to find $\pa_t \gh$. This is a crude estimation, but we do not expect this to cause qualitatively different behavior. Furthermore, $L_J$ is the length scale of the electric field gradient on the cooling front of the injection, which we assume to be $L_J\simeq 1cm$. The initial current density is set to be $J(0)=1MA/m^2$.

To calculate the self-limiting sink, we still need to estimate the effective fragment cross-section $\sum_N{S_p}$ and the flux tube volume $\gD V$. For $\gD V$, we consider a simple torus with width $d=1mm$, minor radius $a=1m$ and major radius $R=6m$, so that
\bqbl
\lbq{eq:deltaV}
\gD V
=
4\gp^2adR
.\eqbl
For $\sum_N{S_p}$, we consider the fragment radius distribution described by the Statistical Fragmentation Model \cite{ParkDistribution} which shows good agreement with experimental results \cite{Gebhart2020FST}:
\bqbl
\lbq{eq:ParkP}
P\left(r_p\right)
=
\R{r_pK_0\left(\gk_p r_p\right)}{I}
,\quad
I
\equiv
\int_0^\infty{r_pK_0\left(\gk_p r_p\right)dr}
=
\gk_p^{-2}
.\eqbl
Here, $K_0$ is the modified Bessel function of the second kind, and $\gk_p$ is the inverse of the characteristic fragment size. For a given $\gk_p$, the average volume of the fragments is then $\left<V_p\right>=6\gp^2\gk_p^{-3}$ and the average fragment cross-section is $\left<S_p\right>=4\gp\gk_p^{-2}$. Thus the effective self-limiting cross-section is
\bqbl
\sum_N{S_p}
=
\R{d}{L_p}N_p\left<S_p\right>
=
\R{d}{L_p}N_p4\gp\gk_p^{-2}
.\eqbl
Here $L_p$ is the length of the fragment plume and the total fragment number $N_p$, assuming all injected materials are converted into fragments instead of gases and droplets, is determined by
\bqbl
N_p n_p \left<V_p\right>
=
N_p n_p 6\gp^2 \gk_p^{-3}
=
N_0
,\eqbl
where $n_p$ is the atom number density of the solid fragment and $N_0$ is the total injected amount of atoms.
For a given $L_p$, we then have
\bqbl
\lbq{eq:sumSp}
\sum_N{S_p}
=
\R{d}{L_p}\R{2N_0}{3\gp n_p\gk_p^{-1}}
.\eqbl
Thus, for a given injection quantity and plume spreading length, the finer shattered the pellet, the stronger the self-limiting effect. This is because the finer fragments have larger surface-to-volume ratio. It should also be noted that combing Eq.\,(\rfq{eq:SLSink}), Eq.\,(\rfq{eq:deltaV}) and Eq.\,(\rfq{eq:sumSp}), the flux tube width $d$ would cancel, thus ultimately our self-limiting sink does not depend on the its detailed choice, but rather only depends on the longitudinal solid fragment concentration of the plume.
Let $N_0\simeq 2.1\times10^{24}$ deuterium atoms, $L_p\simeq 0.5m$ and estimate the solid deuterium atom density to be $n_p\simeq5.85\times10^{28}/m^3$ \cite{Megaw1936N}. For simplicity, we consider fragments with constant radius over time. This is justified by the continuous flow of fragments moving through the flux tube during the SPI process.

\begin{figure*}
\centering
\noindent
\btbl{cc}
\parbox{2.5in}{
    \includegraphics[scale=0.275]{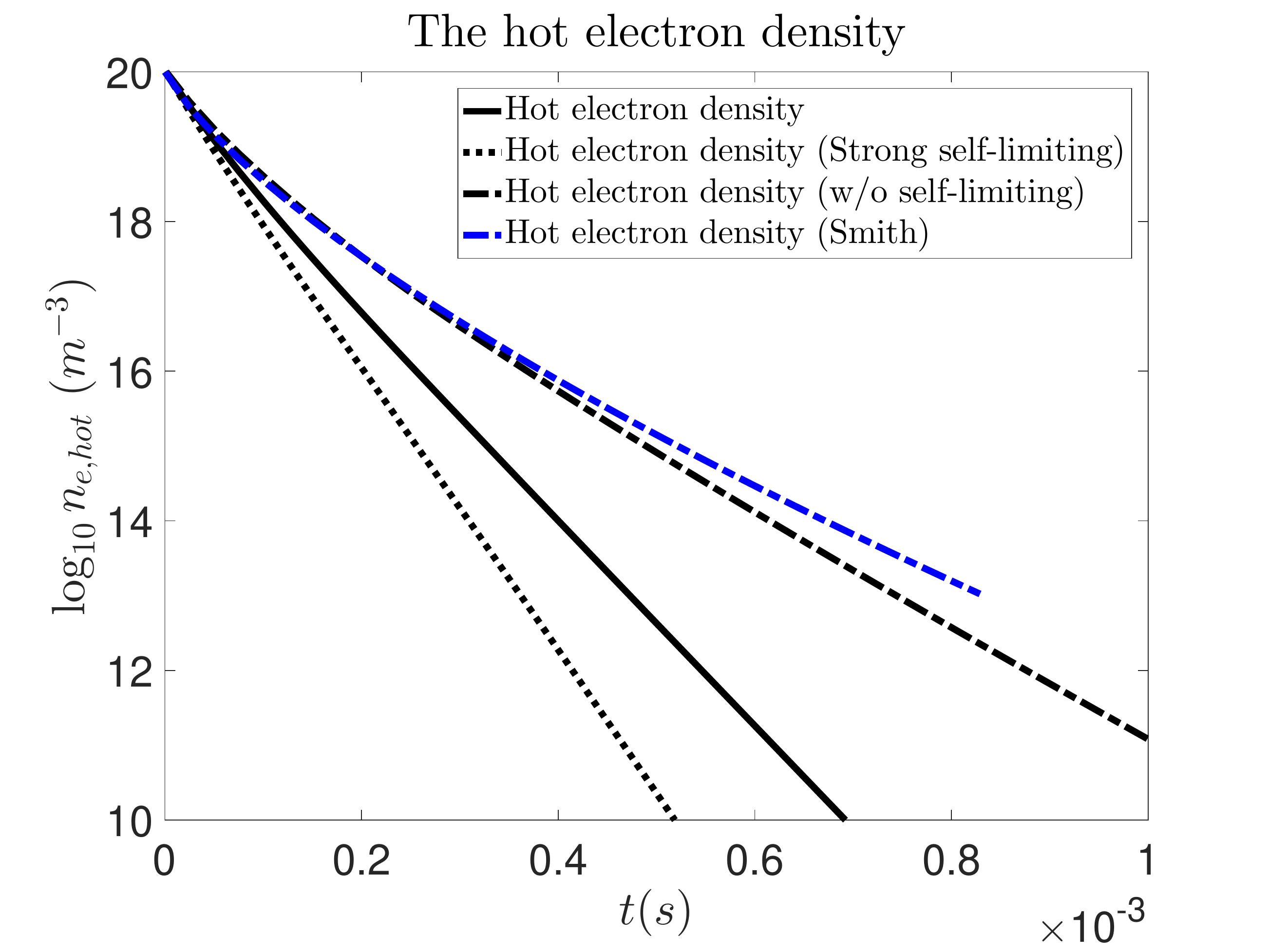}
}
&
\parbox{2.5in}{
	\includegraphics[scale=0.275]{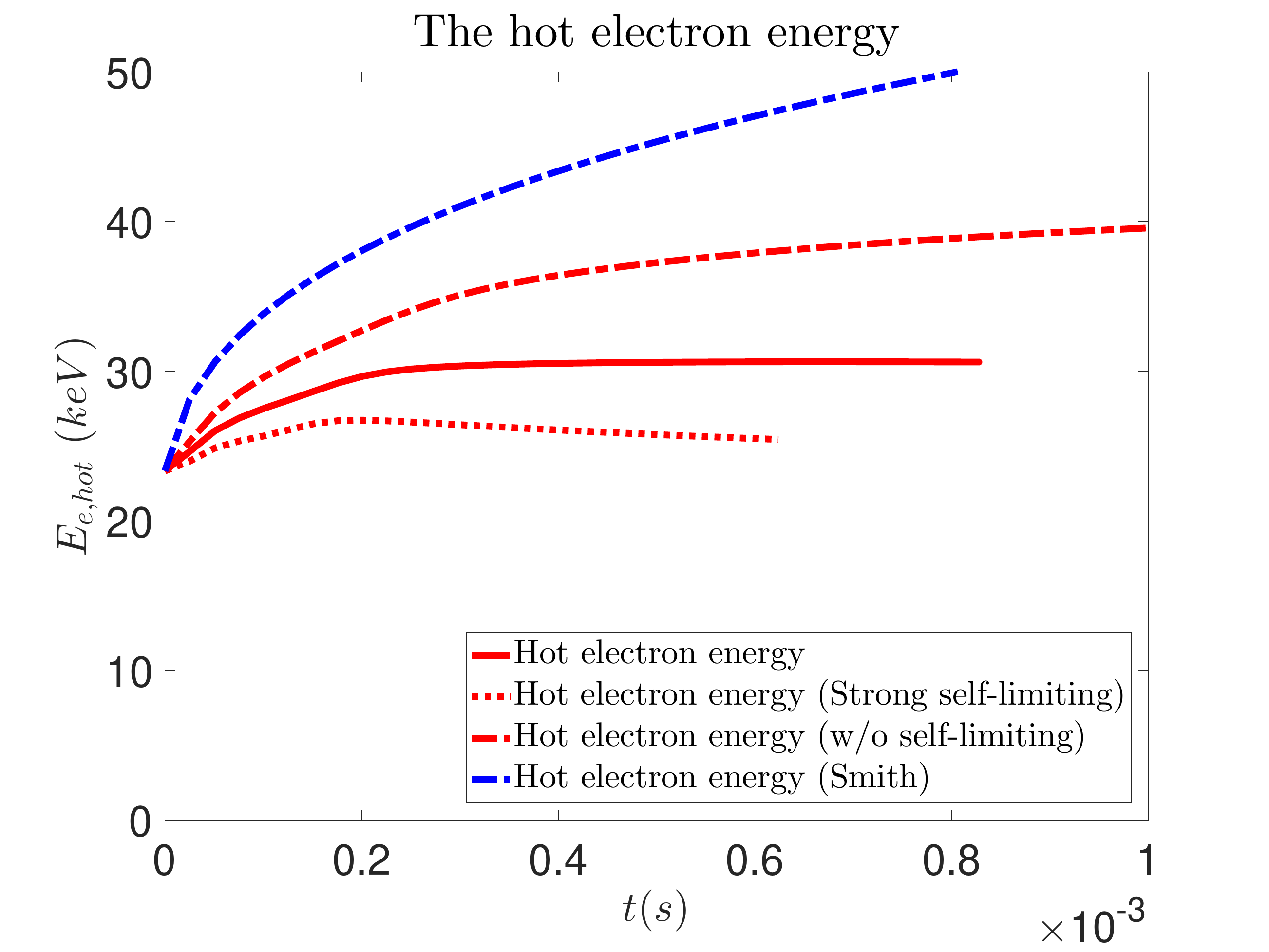}
}
\\
(a)&(b)
\etbl
\caption{(a) The hot electron density and (b) their mean energy as a function of time for $T_{e,hot}(0)=15keV$. The black solid line is the numerical result of hot electron density with $\gk_p^{-1}=0.545mm$, the black dotted line is that of $\gk_p^{-1}=0.268mm$ and the black chained line is that of without self-limiting case. The red solid line is the numerical result of the hot electron mean energy with $\gk_p^{-1}=0.545mm$, the red dotted line is that with $\gk_p^{-1}=0.268mm$, and the red chained line is that of without self-limiting case. The blue chained line in both sub-figures are the analytical Smith result.}
\label{fig:15keVnosink}
\end{figure*}

We first consider the thermalization of hot electrons with initial temperature $T_{e,hot}(0)=15keV$. The comparison of the hot electron density and mean energy evolution are shown in Fig.\,\ref{fig:15keVnosink}. The black lines in Fig.\,\ref{fig:15keVnosink}(a) are the numerical result solving Eq.\,(\rfq{eq:HotEKinetic}), while the blue one the the analytical Smith result according to Eq.\,(\rfq{eq:SmithDensity}). The black solid line is the self-limiting result with $\gk_p^{-1}=0.545mm$, the black dotted line is that with $\gk_p^{-1}=0.268mm$ which we call the ``strong self-limiting'' case, the black chained line is that of the no self-limiting case.
Good agreement can be found between the no self-limiting numerical result and the Smith result until late in time. The minor late time deviation is likely due to the relativistic collision operator produces stronger collision for the high energy electrons compared with the Smith model, thus faster thermalization. On the other hand, the self-limiting cases deviate significantly from the Smith result and shows faster depletion of the hot electron population, with stronger self-limiting sink showing faster hot electron density decay. This is also in consistence with our anticipation. Note that the black dotted line shows such good linearity, suggesting that the hot electron density follows an exponential decay over time. This indicates the self-limiting sink is dominant in the strong self-limiting case. As for the mean energy evolution shown in Fig.\,\ref{fig:15keVnosink}(b), the red solid line the result with $\gk_p^{-1}=0.545mm$, the red dotted line is that with $\gk_p^{-1}=0.268mm$ and the red chained line is that of the no self-limiting case. The blue chained line is the analytical Smith result. It can be seen that without self-limiting effect, the mean hot electron energy rises as would be expected, while the inclusion of the self-limiting effect limits such mean energy rise or even stops it altogether, depending on the strength of the self-limiting sink. This is due to the preferential depletion of the hot population as we discussed above.
The deviation between the red and blue chained line is likely due to the stronger collision produced by our relativistic collision operator as discussed above. The ratio between the Smith result and our numerical result only becomes larger than $1.2$ after $t\simeq 0.5ms$ when the hot electron density is four orders of magnitude smaller than the initial value, hence we consider the effect of this deviation subdominant. This deviation time also coincide with the beginning of density deviation shown in Fig.\,\ref{fig:15keVnosink}(a).

\begin{figure*}
\centering
\noindent
\btbl{c}
\parbox{5.0in}{
    \includegraphics[scale=0.5]{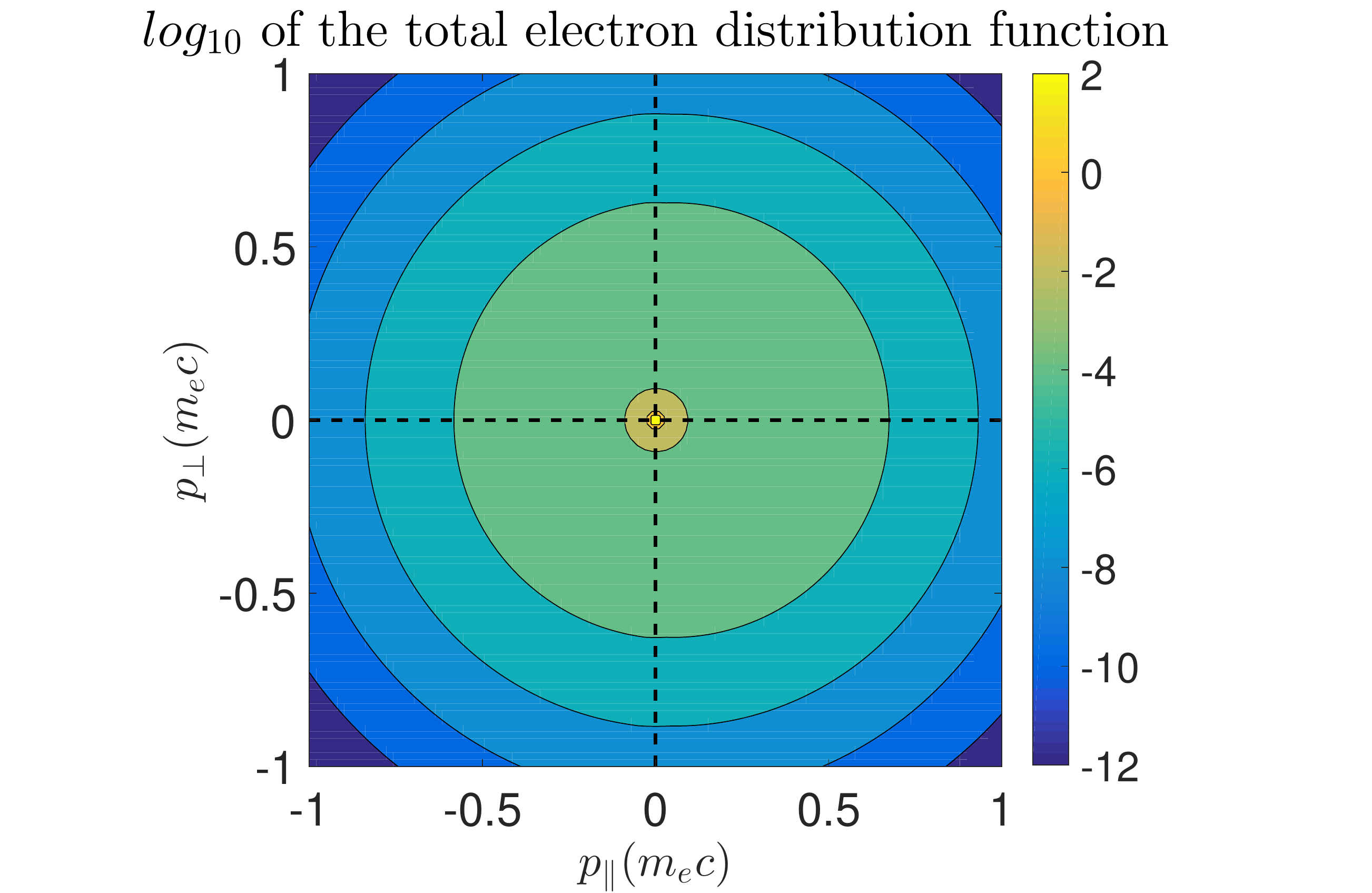}
}
\etbl
\caption{The $\log_{10}$ of the total electron distribution for the no self-limiting and $T_{e,hot}(0)=15keV$ case at $t=0.425ms$ after the beginning of thermalization. The hot tail as well as a slight asymmetric towards to positive $p_\|$ direction can be seen.}
\label{fig:15keVnosink_f}
\end{figure*}

As an example, the hot-tail in the electron distribution function for the no self-limiting case after $t=0.425ms$ since the beginning of the thermalization can be directly seen in Fig.\,\ref{fig:15keVnosink_f}. The isotropic cold electron peak as well as the broad hot-tail electrons can be seen, as well as a slight anisotropy towards the positive $p_\|$ direction due to the electric acceleration, as can be seen by comparing the cold electron population represented by the yellow circle at $(0,0)$ against the outer circles. Such anisotropy is not significant, however, and in the rest of this section we will assume the electron distribution function to be isotropic and take the anisotropy to be a next order effect.

\begin{figure*}
\centering
\noindent
\btbl{cc}
\parbox{2.5in}{
    \includegraphics[scale=0.275]{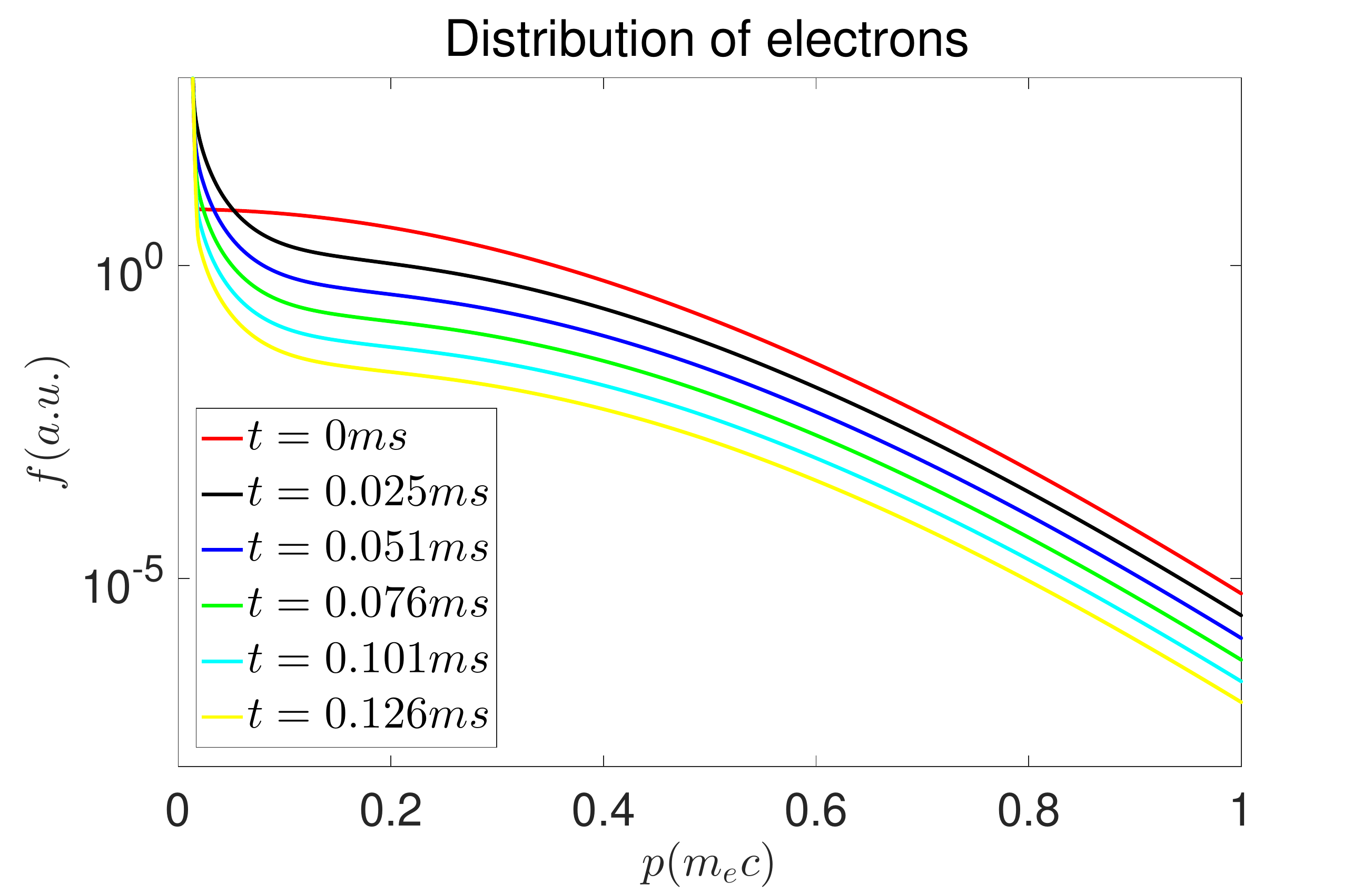}
}
&
\parbox{2.5in}{
	\includegraphics[scale=0.275]{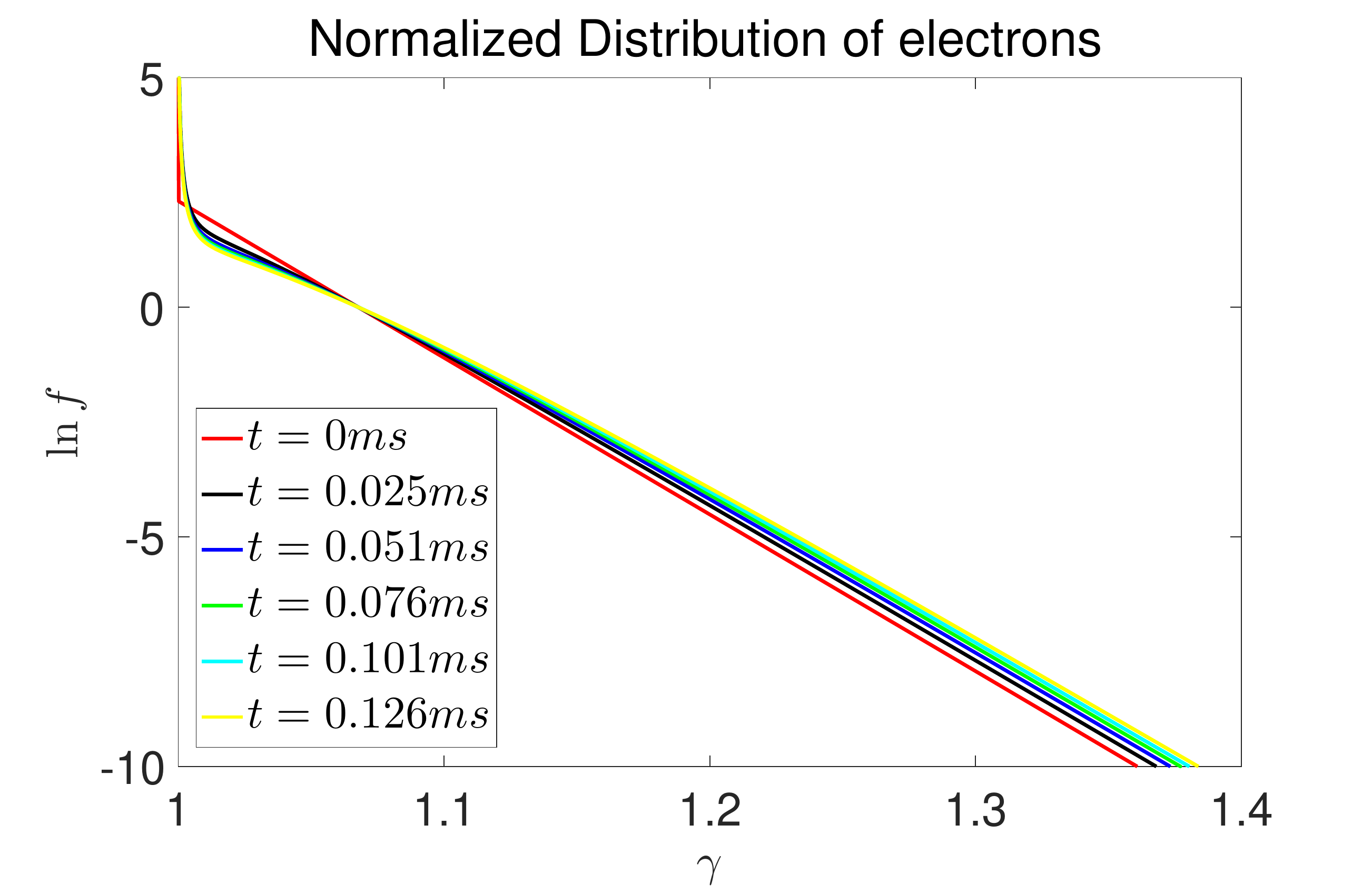}
}
\\
(a)&(b)
\\
\parbox{2.5in}{
    \includegraphics[scale=0.275]{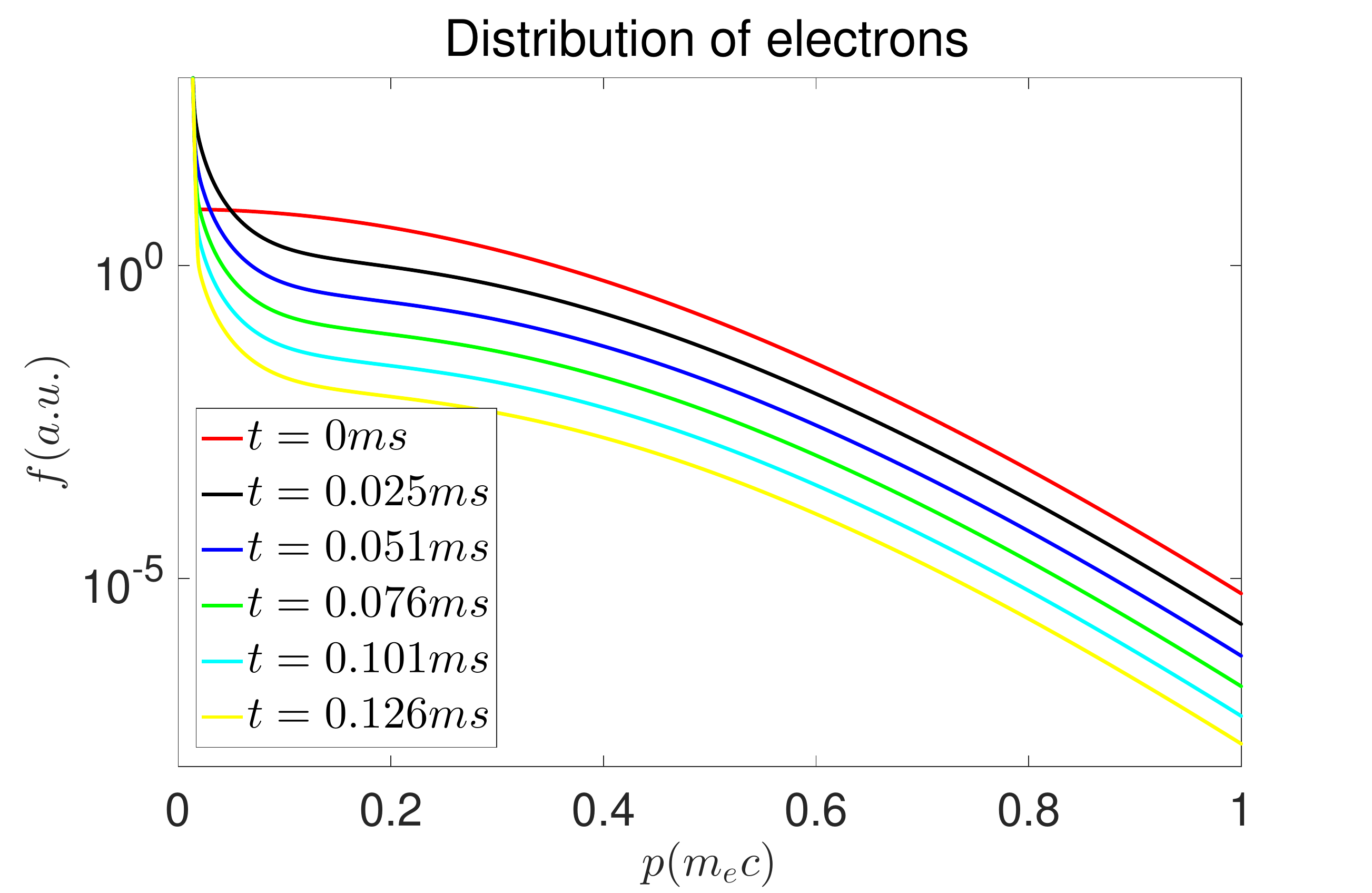}
}
&
\parbox{2.5in}{
	\includegraphics[scale=0.275]{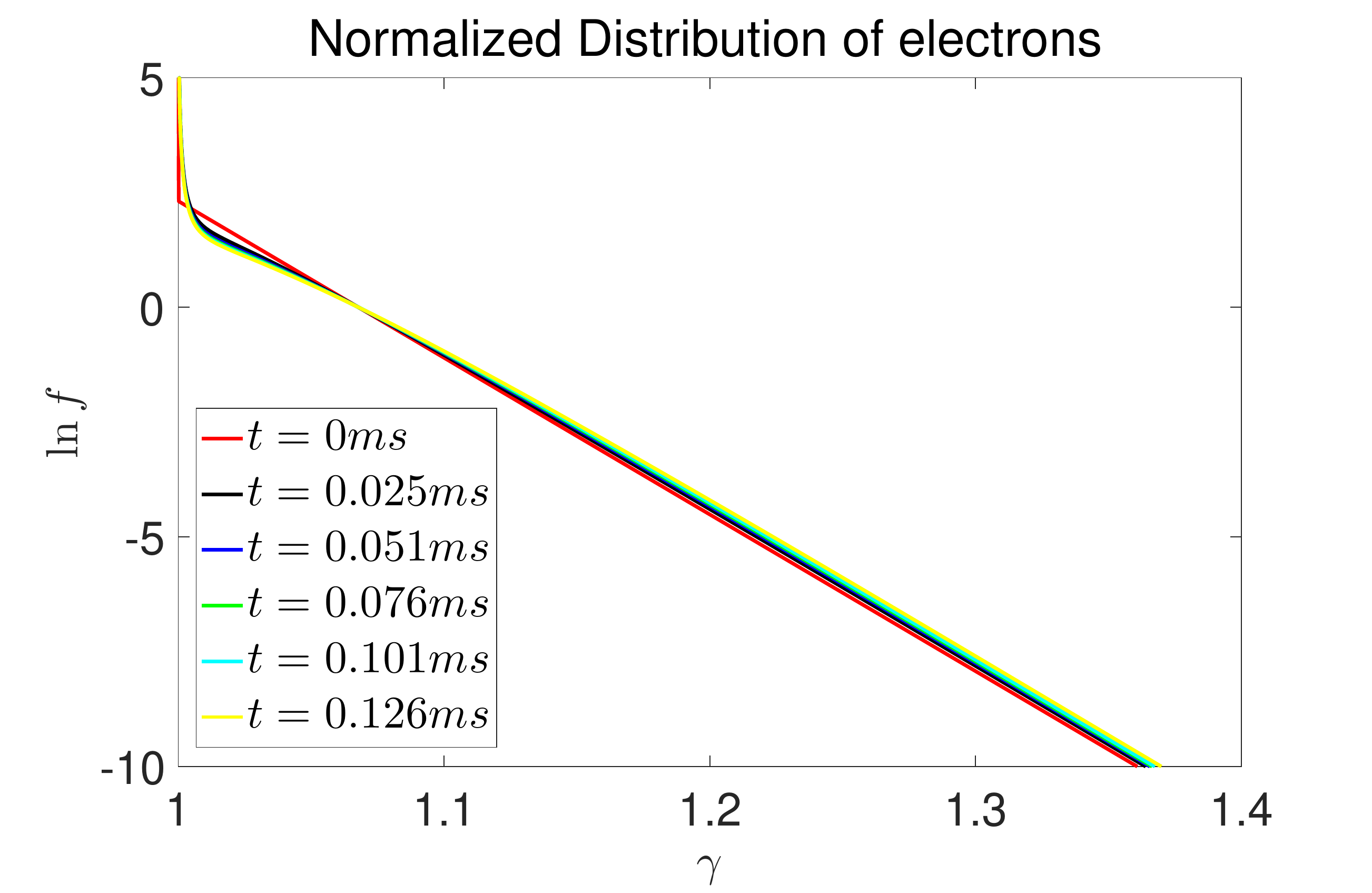}
}
\\
(c)&(d)
\etbl
\caption{(a) The $\gx$ averaged electron distribution with $\gk_p^{-1}=0.545mm$ as a function of momentum $p$, (b) the logarithm of the normalized electron distribution with $\gk_p^{-1}=0.545mm$ as a function of $\gg$, (c) the $\gx$ averaged distribution with $\gk_p^{-1}=0.268mm$ as a function of momentum $p$, (d) the logarithm of the normalized electron distribution with $\gk_p^{-1}=0.268mm$ as a function of $\gg$ at different times for $T_{e,hot}(0)=15keV$.}
\label{fig:15keVf}
\end{figure*}

The time evolution of the $\gx$ averaged electron distribution function, as a function of $p$ and $\gg\equiv\sqrt{1+\left(\R{p}{m_ec}\right)^2}$ respectively, is shown in Fig.\,\ref{fig:15keVf} for the $\gk_p^{-1}=0.545mm$ and $\gk_p^{-1}=0.268mm$ self-limiting case, with distribution at different time represented by the different colored solid lines, and the red solid line represents the initial Maxwellian distribution of the hot electrons. In Fig.\,\ref{fig:15keVf}(a) and (c), the almost self-similar decay of the hot electron distribution is visible as the distribution function maintains more or less the same shape with the initial Maxwellian distribution, despite the rapid decay in its value. In Fig.\,\ref{fig:15keVf}(b) and (d), the logarithm of the normalized distribution as functions of $\gg$ are shown for different times. Here, we choose $\gg$ as the horizontal axis instead of $p$ because the Maxwellian distribution as a function of $\gg$ should be a straight line (the red solid line) in the semi-log plot, which is easy to be compared against.
It is found that only very minor deviation from the initial relativistic Maxwellian distribution. Such self-similar evolution is likely due to the combined effect of the collisional thermalization which preferentially depletes the colder population of a hot electron species, and the self-limiting effect which selectively depletes its hotter population.

\begin{figure*}
\centering
\noindent
\btbl{cc}
\parbox{2.5in}{
    \includegraphics[scale=0.275]{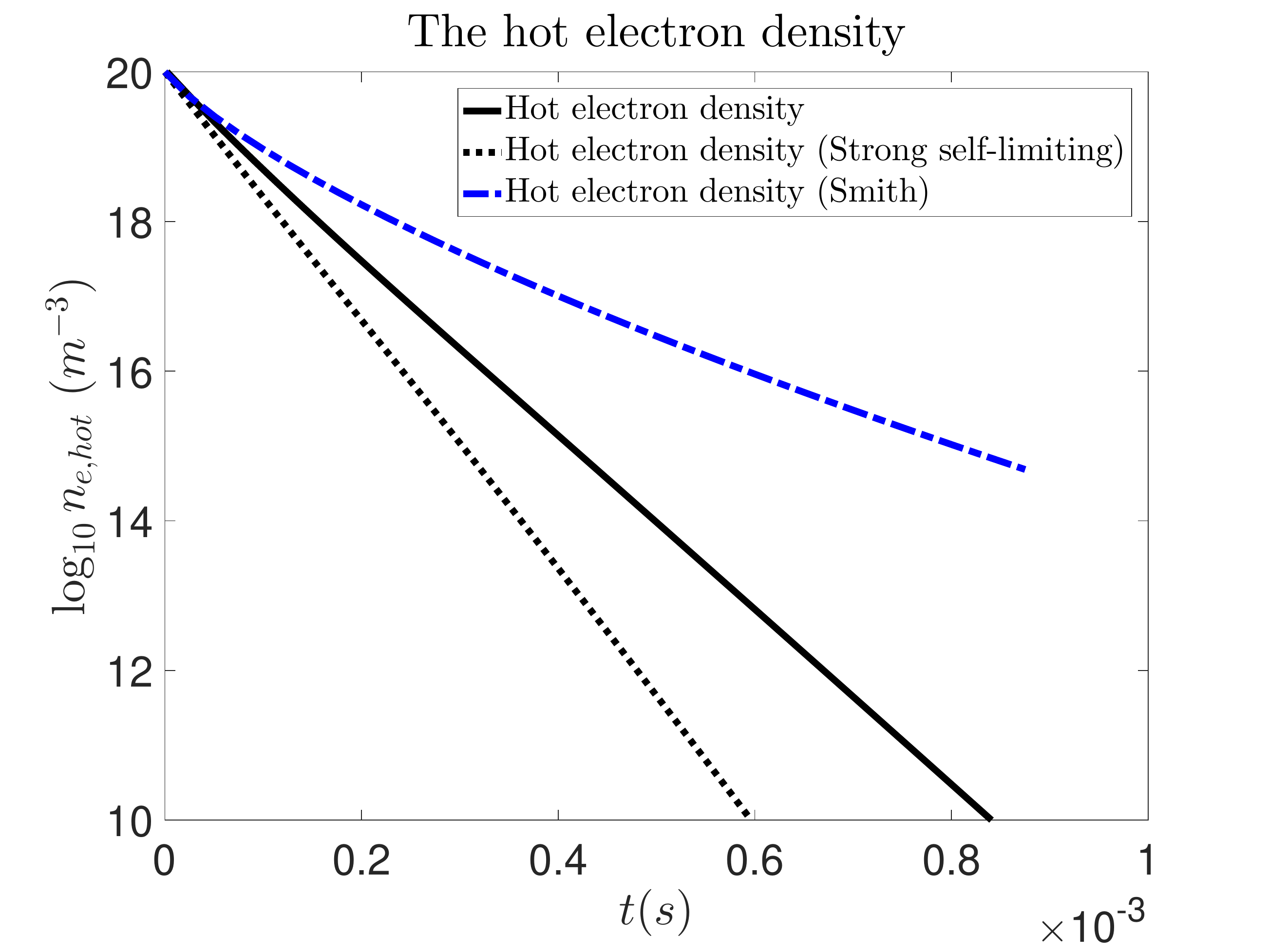}
}
&
\parbox{2.5in}{
	\includegraphics[scale=0.275]{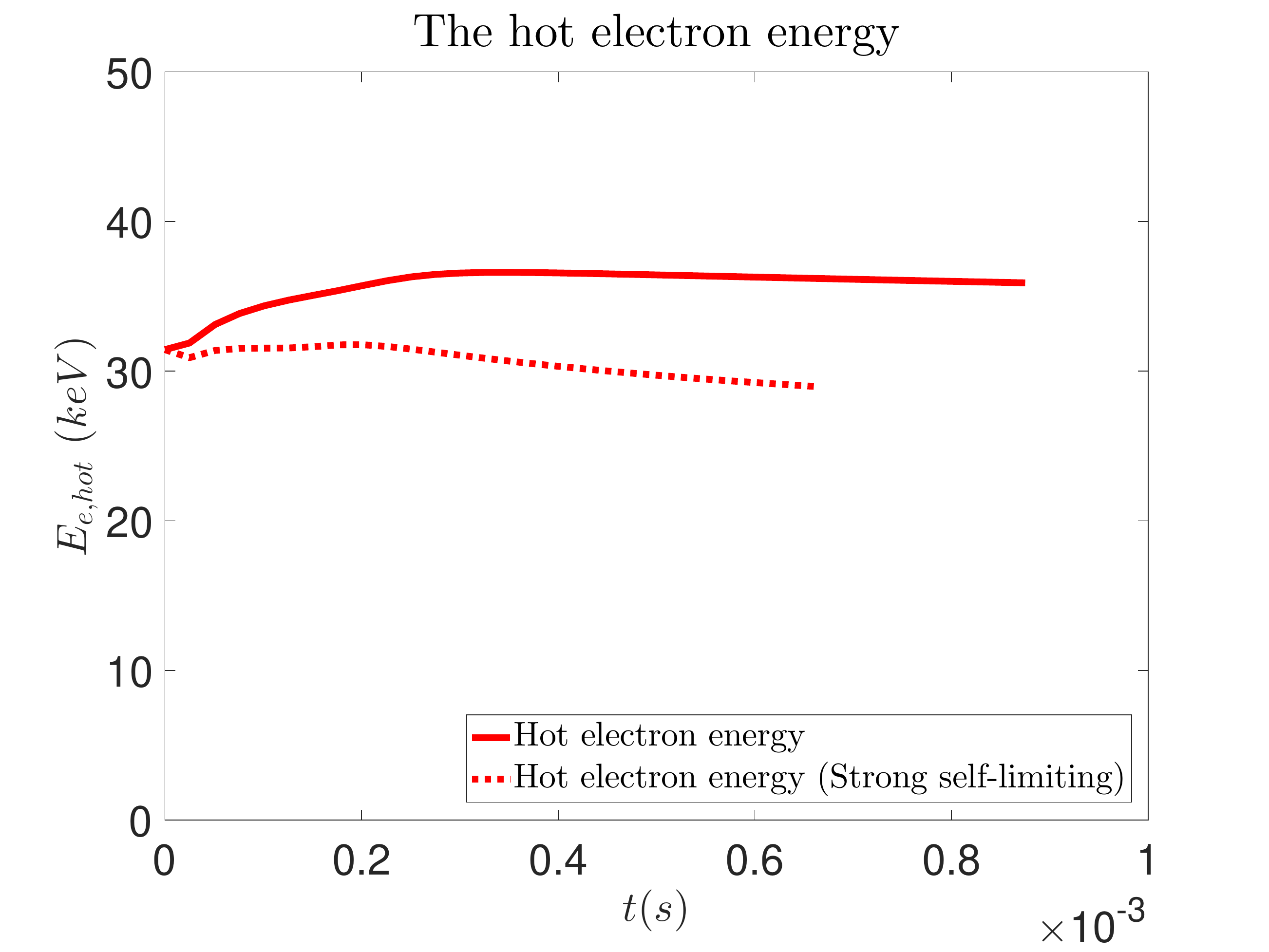}
}
\\
(a)&(b)
\etbl
\caption{(a) The hot electron density and (b) their mean energy as a function of time for $T_{e,hot}(0)=20keV$. The black solid line is the numerical result of hot electron density with $\gk_p^{-1}=0.545mm$ and the black dotted line is that of $\gk_p^{-1}=0.268mm$, the blue chained line is the analytical Smith result. The red solid line is the numerical result of the hot electron mean energy with $\gk_p^{-1}=0.545mm$ and the red dotted line is that with $\gk_p^{-1}=0.268mm$.}
\label{fig:20keV}
\end{figure*}

Similar behavior can be seen for higher initial electron temperature. Here we choose the maximum expected electron temperature in ITER H-mode plasma $T_{e,hot}(0)=20keV$ as an example. The hot electron density and mean energy are once again shown in Fig.\,\ref{fig:20keV}. The black solid line and dotted line in Fig.\,\ref{fig:20keV}(a) represent the hot electron density with $\gk_p^{-1}=0.545mm$ and $\gk_p^{-1}=0.268mm$ respectively, while the blue chained line is the Smith result. Significant deviation from the Smith result is shown even early in the thermalization, and both the solid and dotted black lines show good linearity. This is an indication of the self-limiting sink being dominant. The $\gk_p^{-1}=0.268mm$ also shows faster hot electron depletion compared with the $\gk_p^{-1}=0.545mm$, as would be expected. Fig.\,\ref{fig:20keV}(b) shows similar trend with Fig.\,\ref{fig:15keVnosink}(b), with only limited mean energy rise for the $\gk_p^{-1}=0.545mm$ case and even slightly decreasing mean energy for the $\gk_p^{-1}=0.268mm$ case, as the high energy population is being curtailed by the self-limiting effect.

\begin{figure*}
\centering
\noindent
\btbl{cc}
\parbox{2.5in}{
    \includegraphics[scale=0.275]{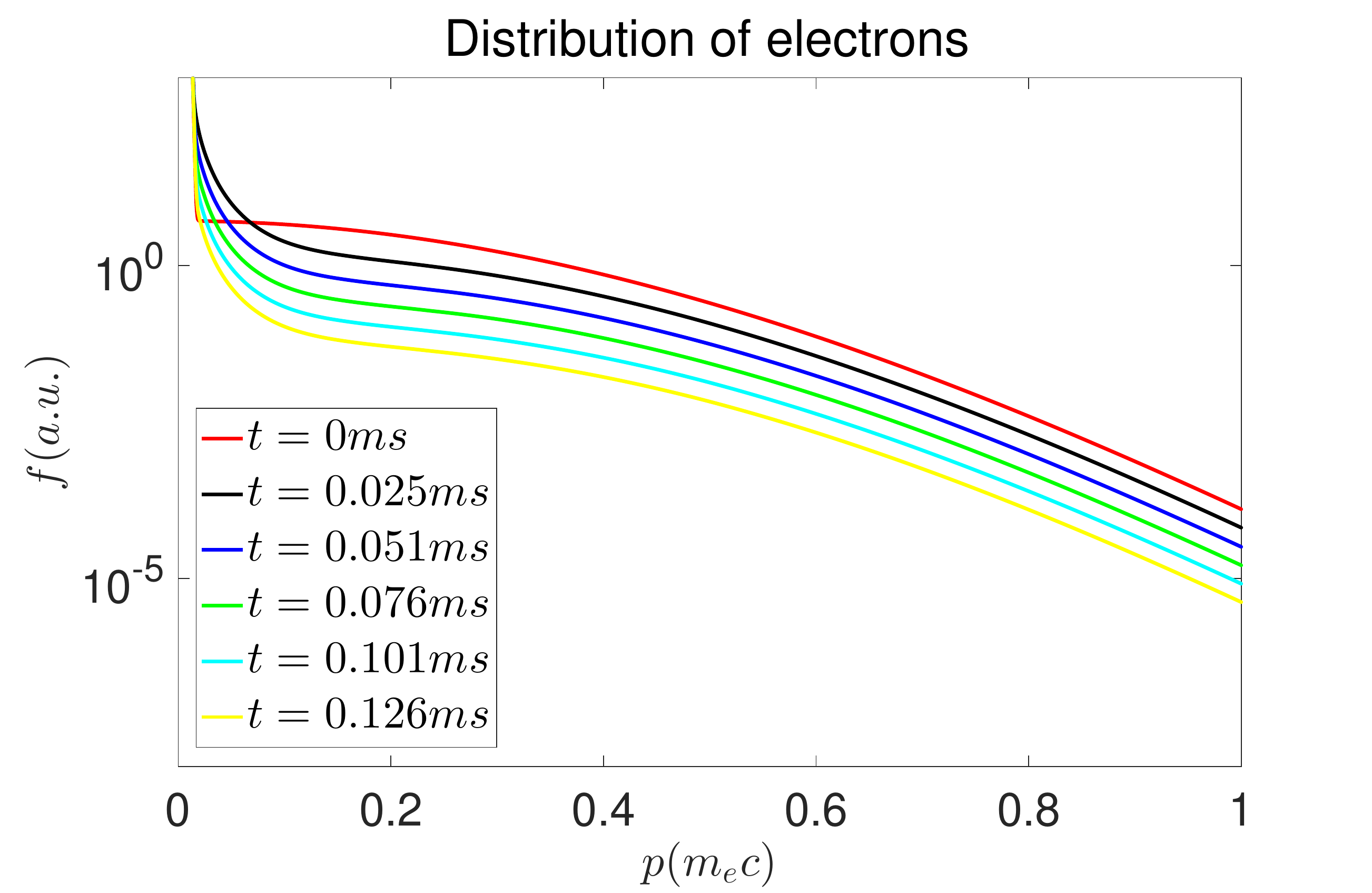}
}
&
\parbox{2.5in}{
	\includegraphics[scale=0.275]{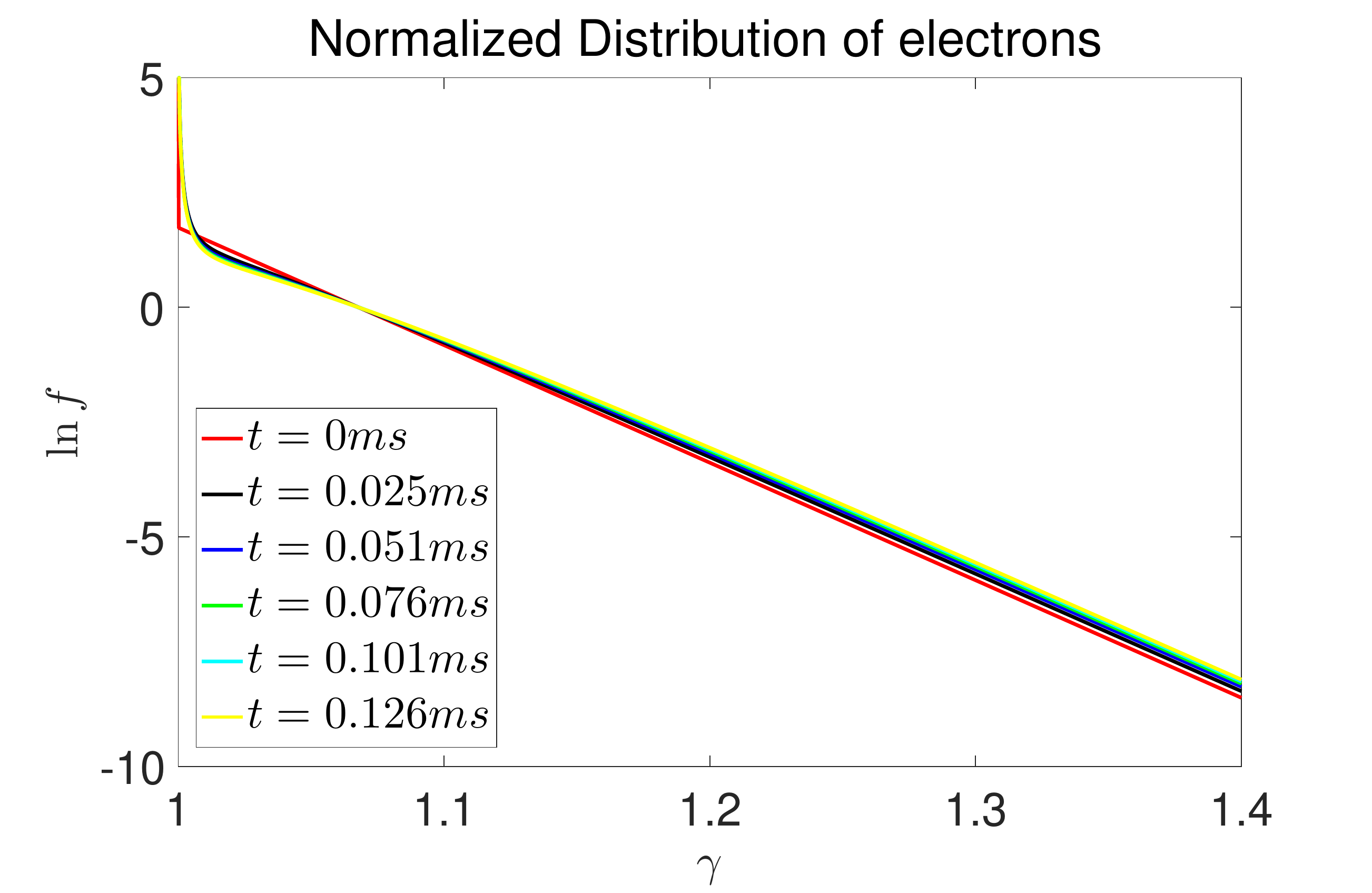}
}
\\
(a)&(b)
\\
\parbox{2.5in}{
    \includegraphics[scale=0.275]{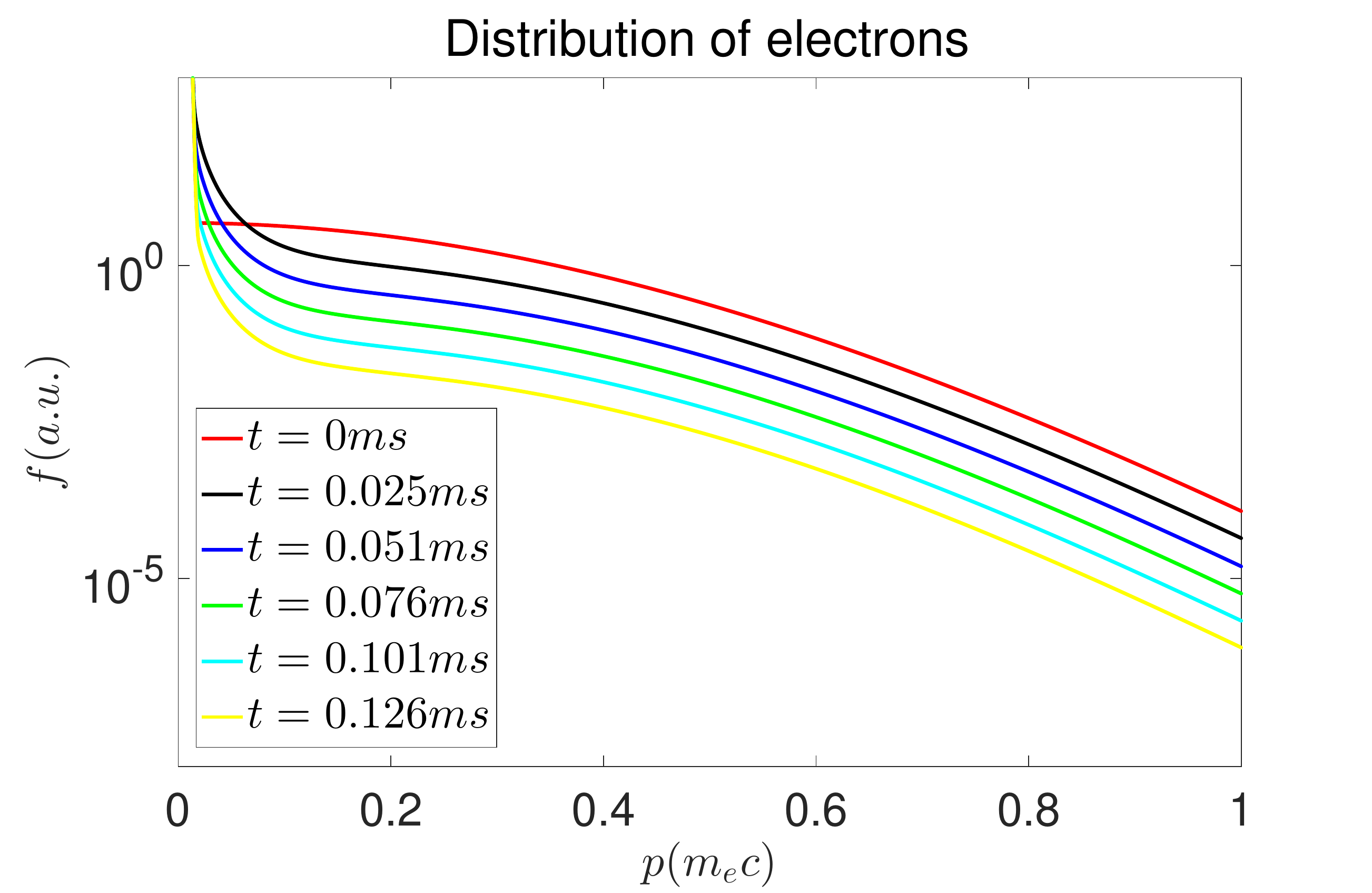}
}
&
\parbox{2.5in}{
	\includegraphics[scale=0.275]{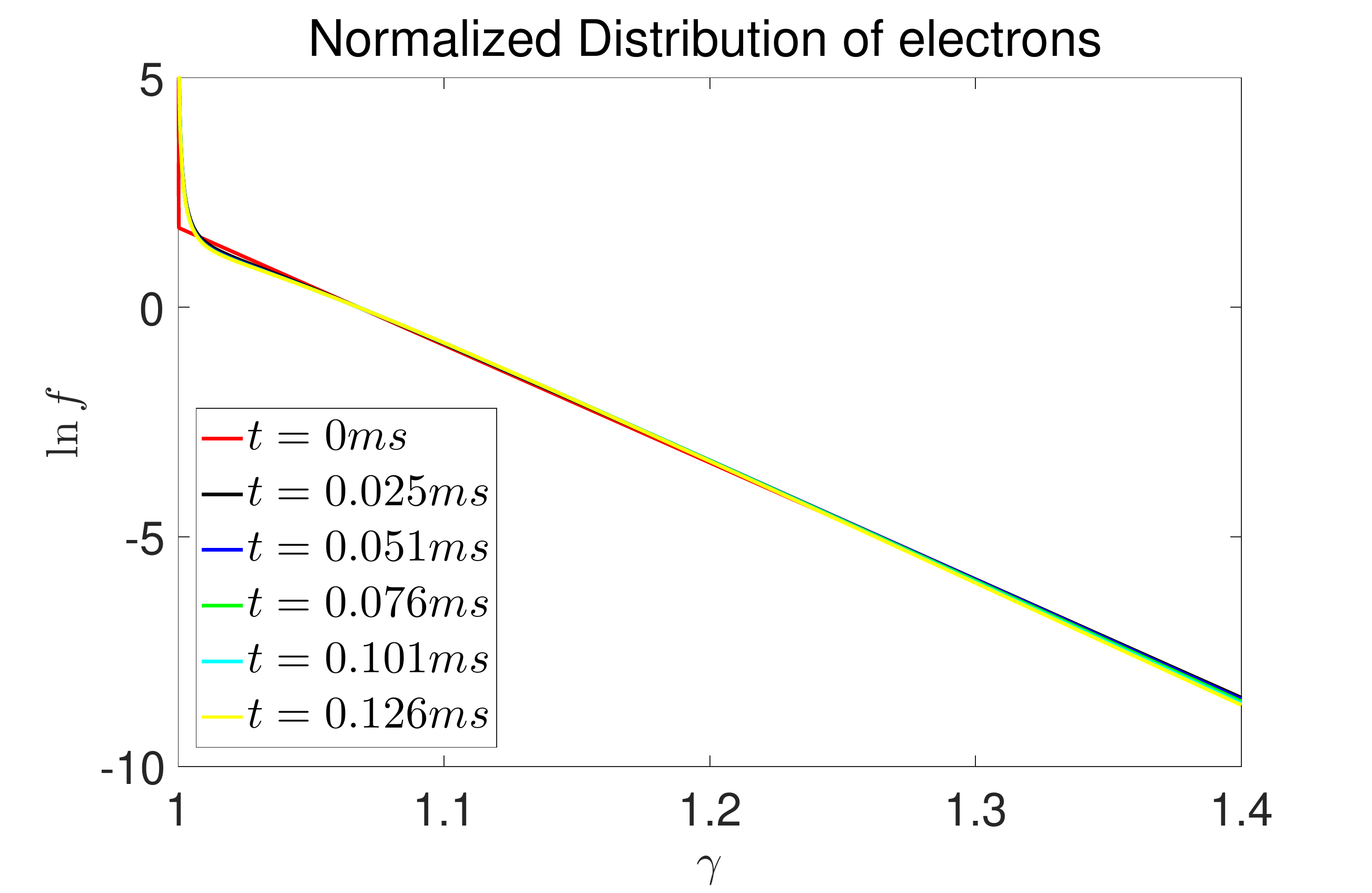}
}
\\
(c)&(d)
\etbl
\caption{(a) The $\gx$ averaged electron distribution with $\gk_p^{-1}=0.545mm$ as a function of momentum $p$, (b) the logarithm of the normalized electron distribution with $\gk_p^{-1}=0.545mm$ as a function of $\gg$, (c) the $\gx$ averaged distribution with $\gk_p^{-1}=0.268mm$ as a function of momentum $p$, (d) the logarithm of the normalized electron distribution with $\gk_p^{-1}=0.268mm$ as a function of $\gg$ at different times for $T_{e,hot}(0)=20keV$.}
\label{fig:20keVf}
\end{figure*}

The self-similar depletion of hot electron distribution also exist in the $T_{e,hot}(0)=20keV$ case as is shown in Fig.\,\ref{fig:20keVf}. Like in the previous case, the electron distribution as a function of $p$ for $\gk_p^{-1}=0.545mm$ and $\gk_p^{-1}=0.268mm$ are shown in Fig.\,\ref{fig:20keVf}(a) and (c) respectively, while that as a function of $\gg$ are shown in Fig.\,\ref{fig:20keVf}(b) and (d) respectively. The red lines are once again the initial Maxwellian distribution, with other colored lines representing that of different times. Very similar behavior with Fig.\,\ref{fig:15keVf} is found, and the hot electron distribution only show very minor deviation from the Maxwellian shape.

\begin{figure*}
\centering
\noindent
\btbl{cc}
\parbox{2.5in}{
    \includegraphics[scale=0.275]{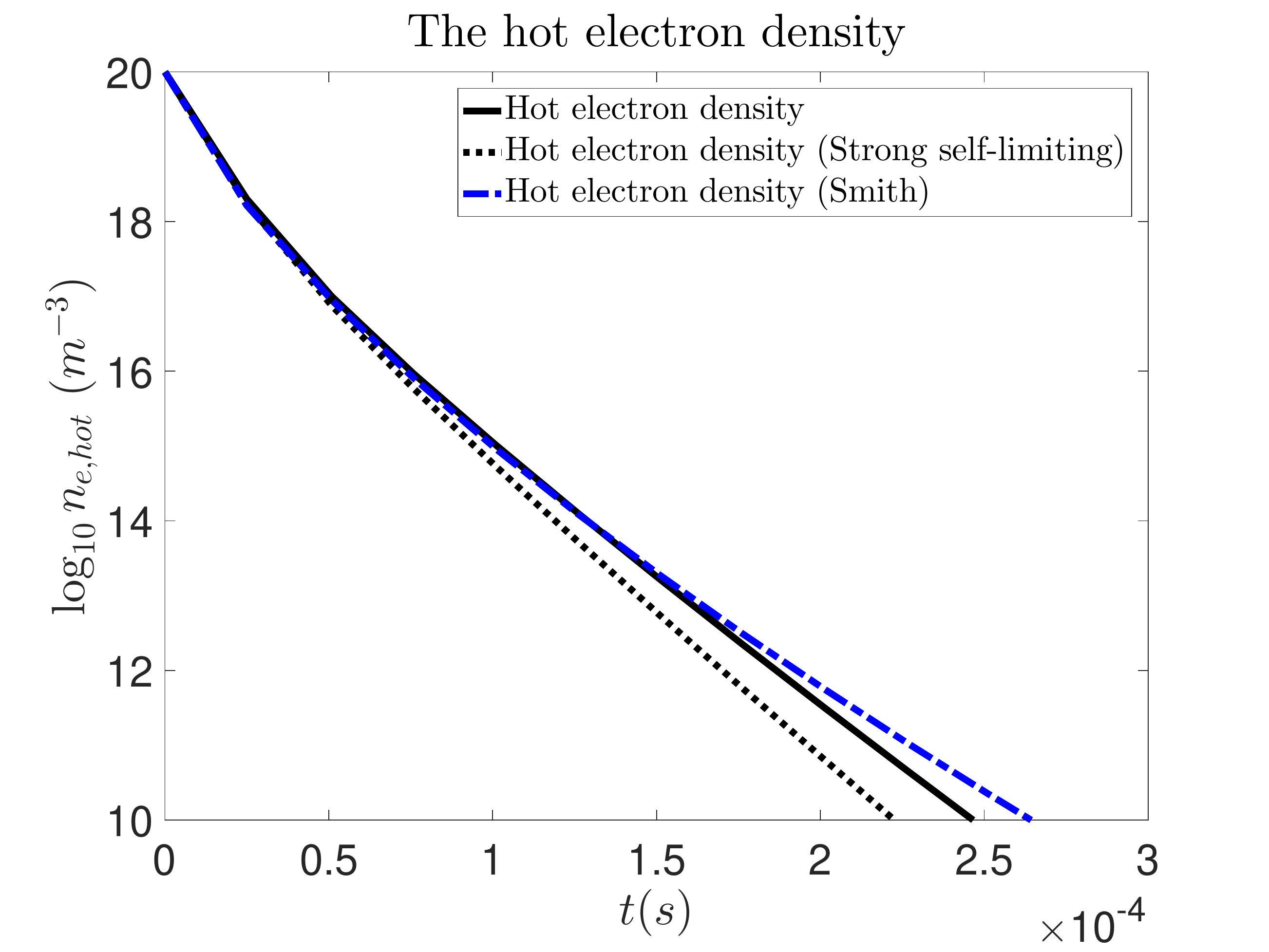}
}
&
\parbox{2.5in}{
	\includegraphics[scale=0.275]{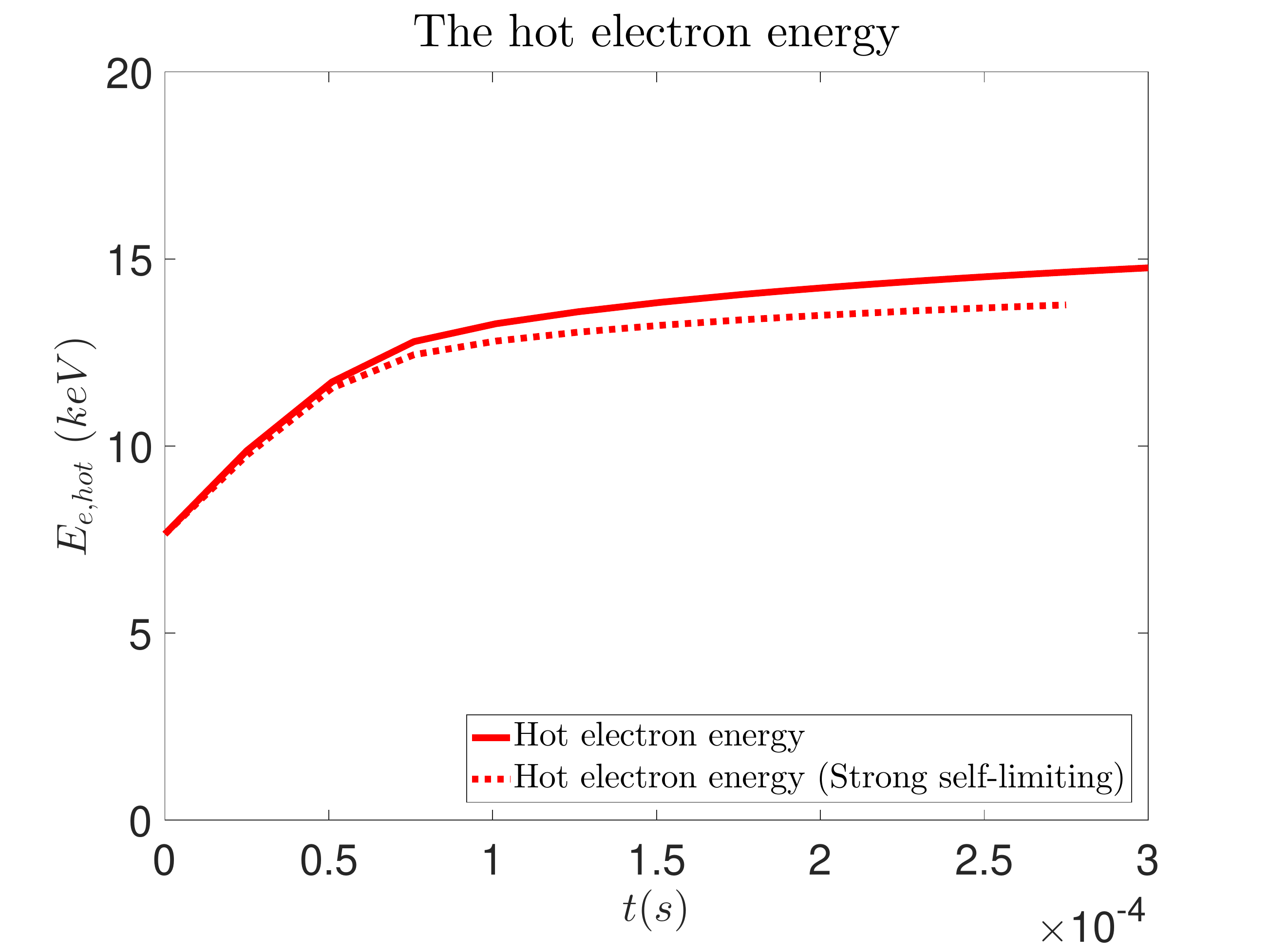}
}
\\
(a)&(b)
\etbl
\caption{(a) The hot electron density and (b) their mean energy as a function of time for $T_{e,hot}(0)=5keV$. The black solid line is the numerical result of hot electron density with $\gk_p^{-1}=0.545mm$ and the black dotted line is that of $\gk_p^{-1}=0.268mm$, the blue chained line is the analytical Smith result. The red solid line is the numerical result of the hot electron mean energy with $\gk_p^{-1}=0.545mm$ and the red dotted line is that with $\gk_p^{-1}=0.268mm$.}
\label{fig:5keV}
\end{figure*}

Such self-similar behavior is not always the case, however, especially when the initial hot electron temperature is not high enough. The evolution of the hot electron density and their mean energy for $T_{e,hot}(0)=5keV$ is shown in Fig.\,\ref{fig:5keV}. Again, in Fig.\,\ref{fig:5keV}(a), the black solid line shows the hot electron density evolution with $\gk_p^{-1}=0.545mm$ self-limiting sink, while the black dotted line show that with $\gk_p^{-1}=0.268mm$ self-limiting sink. The blue chained line shows the Smith result. There is very little difference between the $\gk_p^{-1}=0.545mm$ case with the Smith result, suggesting the thermalization is dominated by collisional thermalization due to the colder initial temperature. The $\gk_p^{-1}=0.268mm$ case shows larger deviation, but only in the late phase when the hot electron mean energy is already significantly raised such deviation becomes apparent, as is shown in Fig.\,\ref{fig:5keV}(b). It should be noted that the density depletion is actually much faster in the $T_{e,hot}(0)=5keV$ due to the strong collision of the low initial temperature plasma, as can be seen comparing Fig.\,\ref{fig:5keV}(a) with Fig.\,\ref{fig:15keVnosink}(a) and Fig.\,\ref{fig:20keV}(a).
The red solid and dotted lines in Fig.\,\ref{fig:5keV}(b) represent the mean energy evolution for $\gk_p^{-1}=0.545mm$ and $\gk_p^{-1}=0.268mm$ respectively. Significant rise of mean energy exists, as opposed to the previous cases shown in Fig.\,\ref{fig:15keVnosink}(b) and Fig.\,\ref{fig:20keV}(b), in accordance with our previous observation that the self-limiting effect is subdominant in the $T_{e_hot}(0)=5keV$ case.

\begin{figure*}
\centering
\noindent
\btbl{cc}
\parbox{2.5in}{
    \includegraphics[scale=0.275]{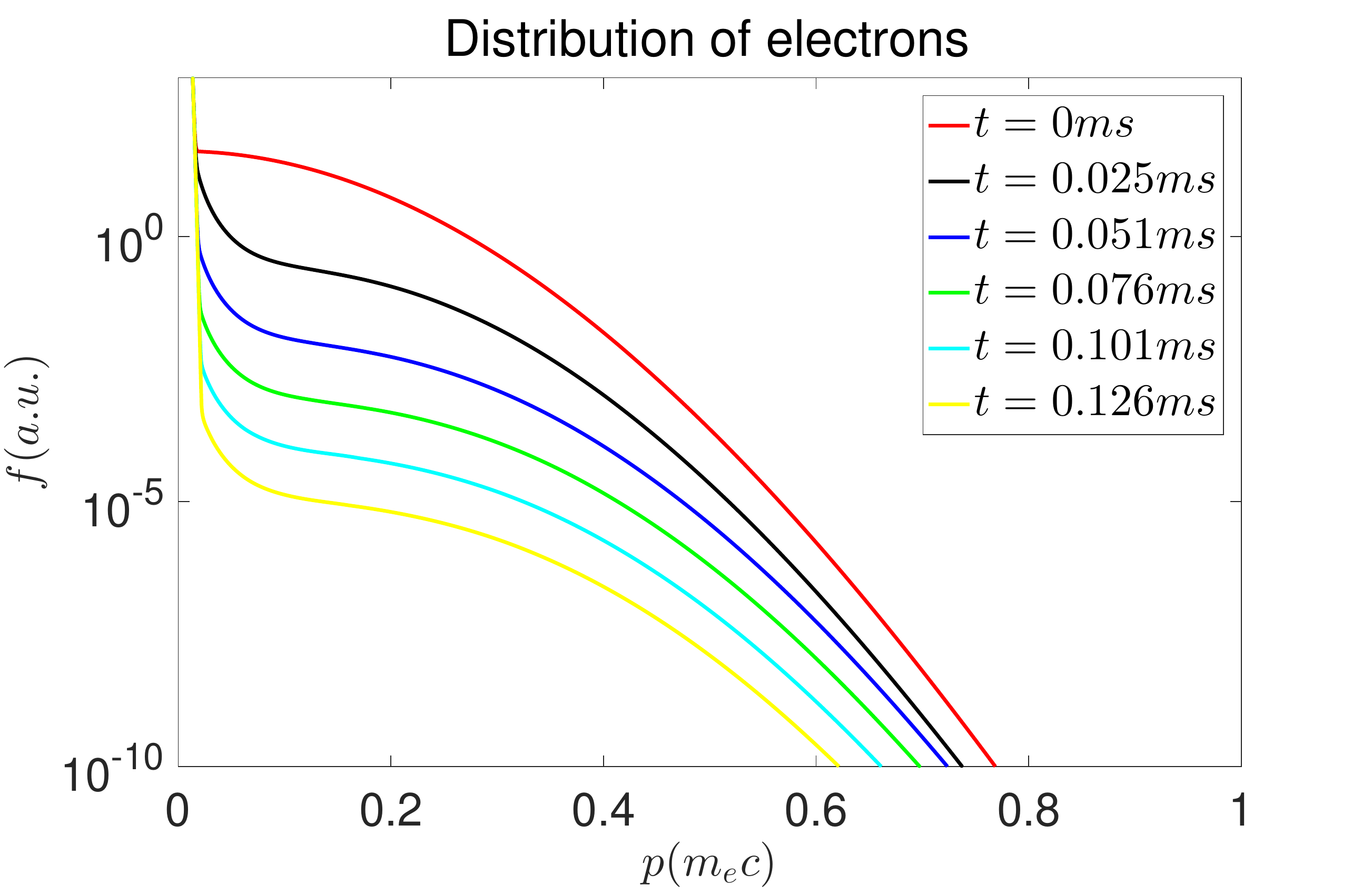}
}
&
\parbox{2.5in}{
	\includegraphics[scale=0.275]{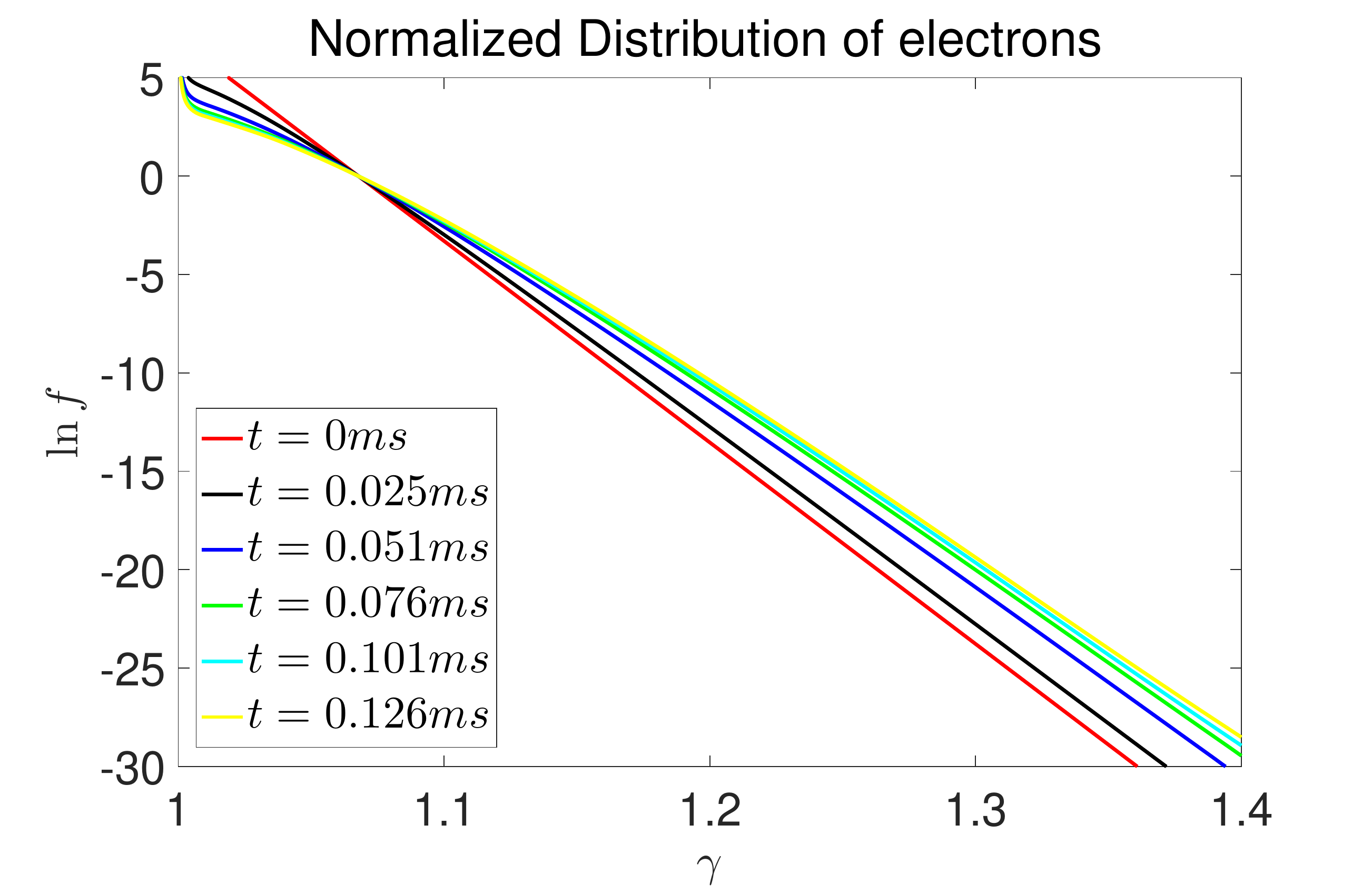}
}
\\
(a)&(b)
\\
\parbox{2.5in}{
    \includegraphics[scale=0.275]{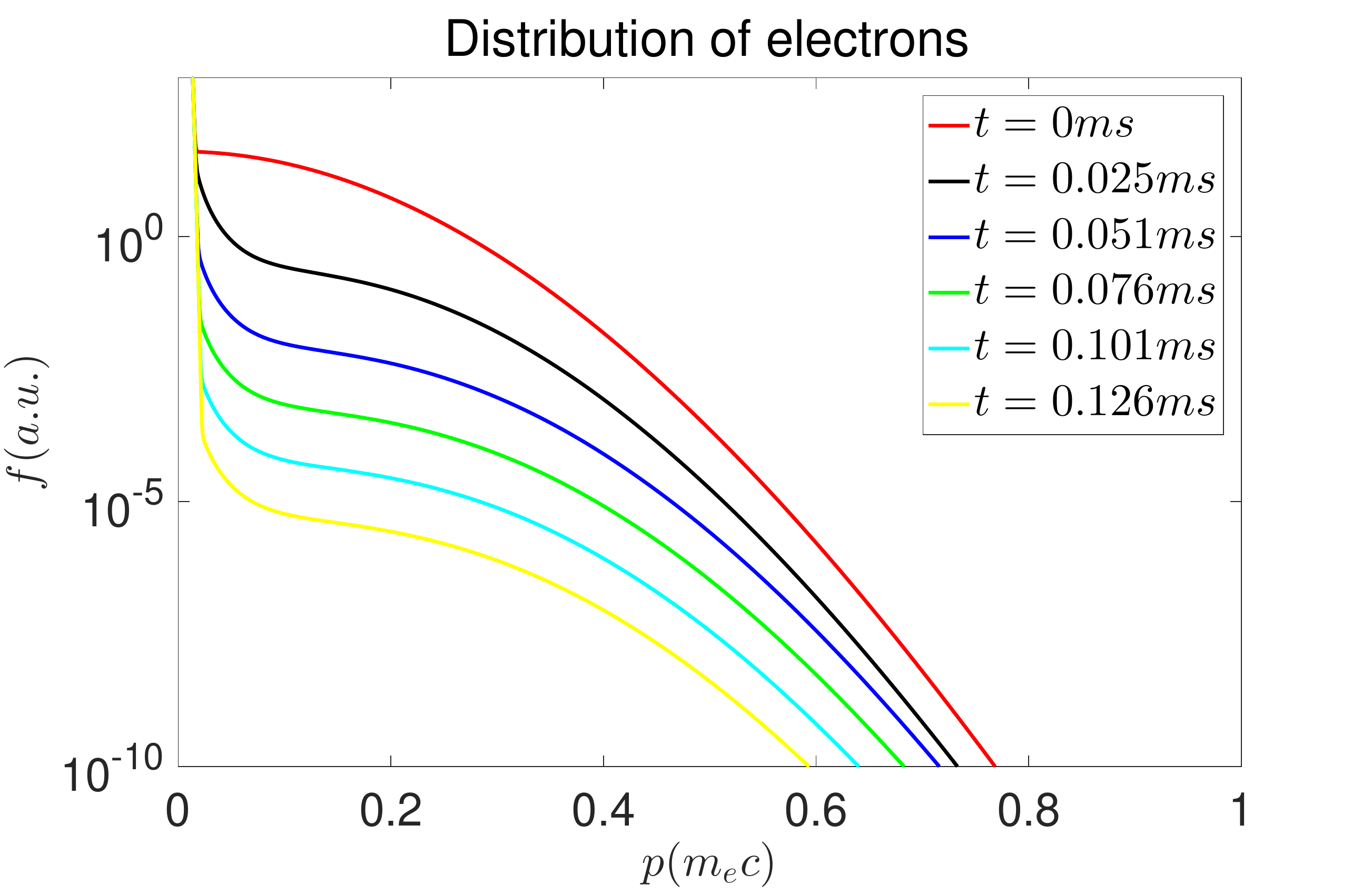}
}
&
\parbox{2.5in}{
	\includegraphics[scale=0.275]{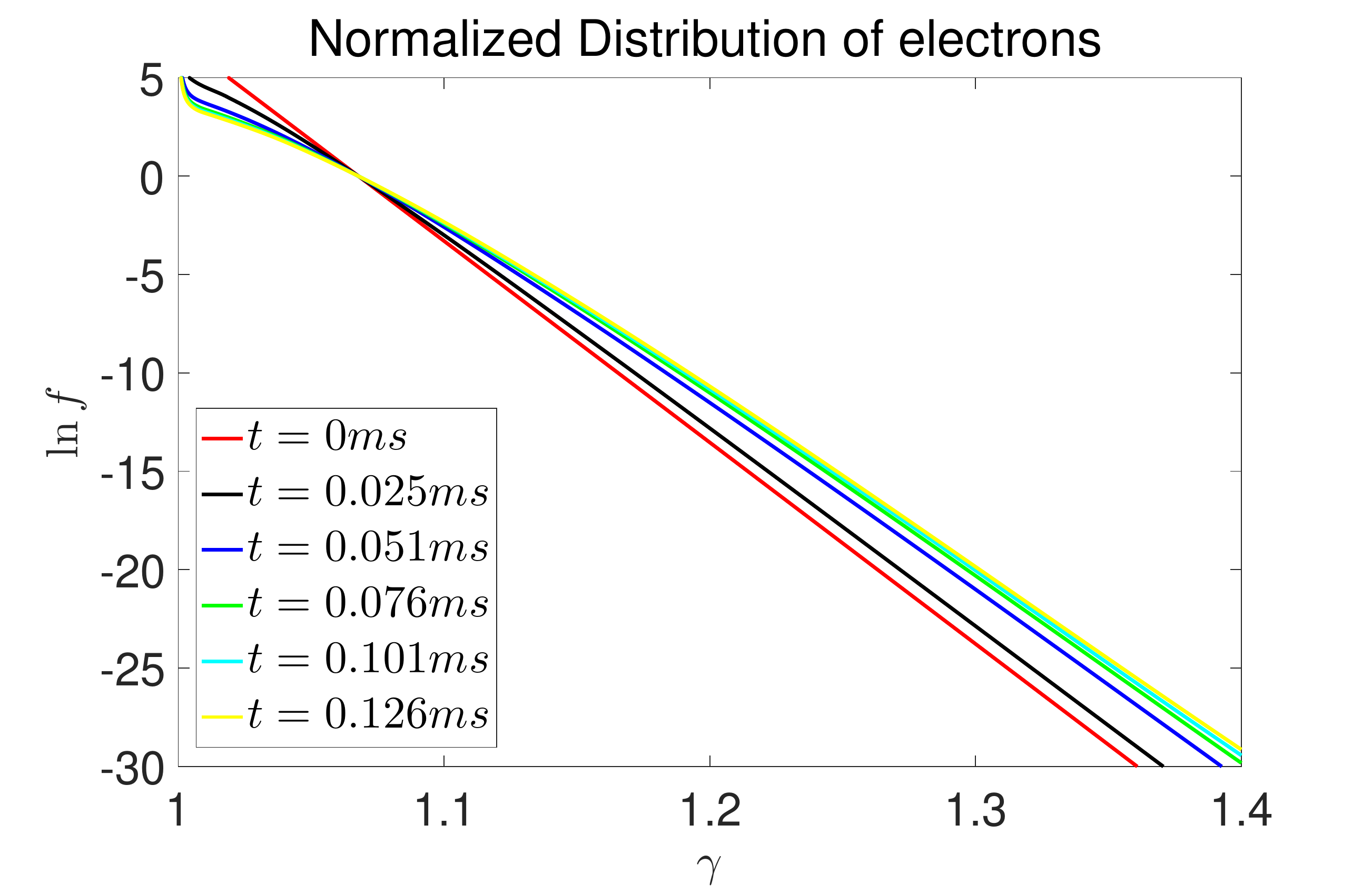}
}
\\
(c)&(d)
\etbl
\caption{(a) The $\gx$ averaged electron distribution with $\gk_p^{-1}=0.545mm$ as a function of momentum $p$, (b) the logarithm of the normalized electron distribution with $\gk_p^{-1}=0.545mm$ as a function of $\gg$, (c) the $\gx$ averaged distribution with $\gk_p^{-1}=0.268mm$ as a function of momentum $p$, (d) the logarithm of the normalized electron distribution with $\gk_p^{-1}=0.268mm$ as a function of $\gg$ at different times for $T_{e,hot}(0)=5keV$.}
\label{fig:5keVf}
\end{figure*}

The subdominant self-limiting strength for this low initial temperature case is also reflected in the distribution function evolution, as is shown in Fig.\,\ref{fig:5keVf}. Again, the red line is the initial Maxwellian distribution, while the other colored lines represent the electron distribution function at different times. Fig.\,\ref{fig:5keVf}(a) shows the distribution as a function of $p$ while Fig.\,\ref{fig:5keVf}(b) shows the normalized one as a function of $\gg$. As opposed to the behaviors shown in Fig.\,\ref{fig:15keVf} and Fig.\,\ref{fig:20keVf}, gradual deviation from the initial Maxwellian shape over time is evident. This is again due to the subdominant role of the self-limiting effect so that the colder population of the hot electrons are preferentially depleted.

Overall, under the combined collisional and self-limiting effect, the thermalization of high initial temperature hot electrons in the bulk plasma could be approximated as the self-similar decay of the hot electron distribution function, while its shape does not deviate much from that of the Maxwellian. The mean hot electron energy shows only limited rise or even slight decrease due to the self-limiting effect. For low initial temperature on the order of $5keV$, however, the self-limiting effect is subdominant, and the thermalization is mostly determined by the collisional process with good agreement with the Smith result. On the other hand, the lower initial temperature case also experiences shorter thermalization time, reducing the finite thermalization time effect on the ablation rate. Thus we will disregard the lower initial temperature behavior and model the hot electron thermalization using the higher temperature behavior in the following sections.

\subsection{Analytical modelling of hot tail electron depletion for fluid coupling}
\label{ss:AnalyticalThreshold}

Since, with high enough initial electron temperature, the hot electron distribution does not deviate much from that of the Maxwellian, we could directly use existing power-law \cite{Zhang2020NF} to estimate their contribution to the ablation rate.
To do this, it is desirable to model the above studied hot-tail depletion in a fluid simulation where the ablation rate, the fragment size, the bulk plasma temperature and density are self-consistently evolved. Direct coupling between the kinetic simulation and the fluid simulation could be costly, however, and we therefore construct an approximate model in this subsection instead.

We model the hot-tail thermalization as the thermalization between two fluid species, the low density hot electron species and the high density bulk cold electron species. In accordance with numerical observations of high initial temperature electron thermalization in Section \ref{ss:NumericalThermalization}, we assume the hot electron mean energy, thus their temperature, is constant in presence of the self-limiting effect. Meanwhile, the hot-tail electron density experiences both the collisional sink and the self-limiting sink. As a first attempt, we hereby neglect the important question of stochastic transport of the hot-tail electrons during the injection process.

Without the self-limiting effect, the hot electron density depletion could be obtained by taking the time derivative of Eq.\,(\rfq{eq:SmithDensity}):
\bqbl
\lbq{eq:Smith_dndt}
\left(\R{\pa}{\pa t}n_{e,hot}\right)_c
&
=
&
-\R{\pa s}{\pa t}\R{\pa S}{\pa s}\R{4e^{-S^{2/3}}}{3\sqrt{\gp}}n_{e,hot}(0)
\nonumber
\\
&
=
&
-\R{e^4n_{e,cold}\ln{\gL}}{\gp\ge_0^2m_e^2c^3}\R{1}{2S^{1/3}+e^{S^{2/3}}\sqrt{\gp}erfc(S^{1/3})}n_{e,hot}(t)
\nonumber
\\
&
=
&
-\R{1}{2\sqrt{2}}\R{e^4n_{e,cold}\ln{\gL}}{\gp\ge_0^2m_e^2v_0^3}
\R{\gq_0^{-3/2}}{2S^{1/3}+e^{S^{2/3}}\sqrt{\gp}erfc(S^{1/3})}n_{e,hot}(t)
.\eqbl
If the distribution function does not deviate much from that of the Maxwellian, Eq.\,(\rfq{eq:Smith_dndt}) would actually underestimate the collisional sink, hence we consider it as a lower estimation. On the other hand, the self-limiting sink could be expressed as
\bqbl
\left(\R{\pa}{\pa t}n_{e,hot}\right)_{sl}
=
\int_1^{\infty}{\left(\R{\pa f(\gg)}{\pa t}\right)_{sl}d\gg}
.\eqbl
Based on our observation of the high initial temperature hot electrons in Section \ref{ss:NumericalThermalization}, we let the distribution function to have the following Maxwell-J\"{u}ttner form:
\bqbl
f(\gg)
=
\R{\gg^2\gb}{\gq K_2\left(\gq^{-1}\right)}\exp{\left(-\R{\gg}{\gq}\right)}
.\eqbl
Here $\gb=v/c$. Thus the integrated sink has the following form:
\bqbl
\lbq{eq:SL_dndt}
\left(\R{\pa}{\pa t}n_{e,hot}\right)_{sl}
&
=
&
-\R{c}{2}n_{e,hot}\left(t\right)\R{\sum_N S_p}{\gD V} \int_1^{\infty}{\R{\gg^2\gb^2}{\gq K_2\left(\gq^{-1}\right)}\exp{\left(-\R{\gg}{\gq}\right)}d\gg}
\nonumber
\\
&
=
&
-\R{c}{2}n_{e,hot}\left(t\right)\R{\sum_N S_p}{\gD V} \int_1^{\infty}{\R{\gg^2-1}{\gq K_2\left(\gq^{-1}\right)}\exp{\left(-\R{\gg}{\gq}\right)}d\gg}
\nonumber
\\
&
=
&
-c\R{\sum_N S_p}{\gD V} \exp{\left(-\gq^{-1}\right)}\R{\gq\left(1+\gq\right)}{K_2\left(\gq^{-1}\right)}n_{e,hot}\left(t\right)
.\eqbl
Here, $K_\gn\left(x\right)$ is the modified Bessel function of the second kind. Since Eq.\,(\rfq{eq:SL_dndt}) is obtained from integration over a relativistic Maxwellian distribution, it is also considered as a lower estimation should the distribution have a Smith-like hot-tail as is in the pure collisional case.

\begin{figure*}
\centering
\noindent
\btbl{c}
\parbox{5.0in}{
    \includegraphics[scale=0.5]{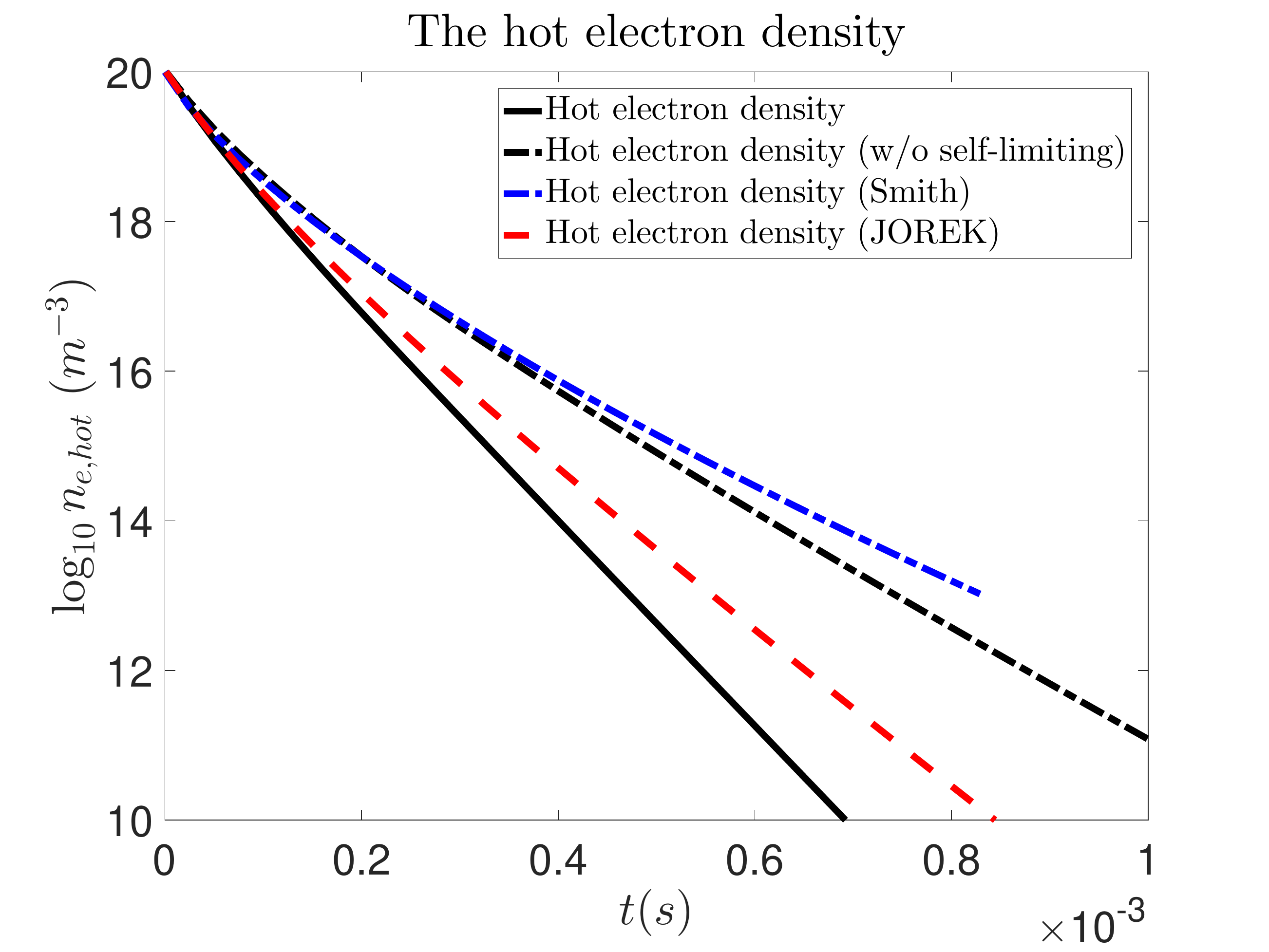}
}
\etbl
\caption{The hot electron density depletion as a function of time for $T_{e,hot}(0)=15keV$. The black solid line is the numerical result of hot electron density with $\gk_p^{-1}=0.545mm$, the black chained line is that of without self-limiting case and the red dashed line is the result from combining Eq.\,(\rfq{eq:Smith_dndt}) and Eq.\,(\rfq{eq:SL_dndt}). Apart from the red dashed line, the other lines are merely copied from that of Fig.\,\ref{fig:15keVnosink}.}
\label{fig:15keV_JOREK}
\end{figure*}

Combining Eq.\,(\rfq{eq:Smith_dndt}) and Eq.\,(\rfq{eq:SL_dndt}) and assuming the hot electron temperature to be constant, we have the governing equations for the hot electron density and temperature evolution in the presence of the self-limiting effect.
As we have mentioned above, both Eq.\,(\rfq{eq:Smith_dndt}) and Eq.\,(\rfq{eq:SL_dndt}) represent lower estimations of the actual hot electron density depletion, so that the thermalization process in reality will be stronger than we presumed in the following sections. An example of such effect can be seen in Fig.\,\ref{fig:15keV_JOREK}, where the combined analytical result is shown in the red dashed line, while the numerical results are copied from Fig.\,\ref{fig:15keVnosink}. Comparing the red dashed line and the numerical result of the black solid line, it can be seen that in the early phase our analytical model produce reasonable agreement, while the analytical model underestimates the density depletion after the density is already several orders of magnitude smaller, as would be expected from the above discussion.

Meanwhile, it is also of interest to look at the case without the self-limiting effect. In this case we will only keep Eq.\,(\rfq{eq:Smith_dndt}) as the hot electron density sink, while the hot electron temperature is no longer constant, but follows a fitted curve of mean energy increase in accordance with the Smith result. As is shown in Fig.\,\ref{fig:15keVnosink}(b), this would likely result in an overestimated mean hot electron energy. However, since the ratio between the Smith result and our numerical result is less than $1.2$ until the hot electron density is already severely depleted, we consider this overestimation as a next order effect.

\section{Overall impact on the injection assimilation and penetration}
\label{s:OverallImpact}

Based on the numerical observation and analytical modelling above regarding the hot electron thermalization, we further study the overall hot-tail impact on the SPI assimilation and penetration of ITER plasmas. In Section \ref{ss:JOREKModelling}, we will introduce the implementation of hot-tail ablation mechanism in JOREK as well as our governing equations. The target equilibrium and the SPI configuration are introduced in Section \ref{ss:Equilibrium}. In Section \ref{ss:JOREKResult}, axisymmetric JOREK simulations of D2 SPI into ITER H-mode equilibrium are performed with different injection configuration. The simulation results with hot-tail effect will be compared to instantaneously thermalized fluid results.

\subsection{JOREK modelling of the finite thermalization time effect}
\label{ss:JOREKModelling}

JOREK is a 3D nonlinear code with both reduced MHD and full MHD capability \cite{Hoelzl2021NF}, although for the purpose of this study we will only utilize its axisymmetric version. For the completeness of this paper, we write down the governing equations we solve:
\bqbl
\lbq{eq:MagField}
\waB
=
F_0\na\gj
+\na\gy\times\na\gj
,\eqbl
\bqbl
\lbq{eq:VelField}
\wsv
=
v_\|\waB
-R^2\na u\times\na \gj
,\eqbl
\bqbl
\lbq{eq:InductionEq}
\R{\pa\gy}{\pa t}
&
=
&
\gh\left(T_e\right)\gD^*\gy
-R\left\{u,\gy\right\}
-F_0\R{\pa u}{\pa\gj}
,\eqbl
\bqbl
\lbq{eq:AmpereEq}
j
&
=
&
\gD^*\gy
\equiv
R^2\na\cdot\left(\R{1}{R^2}\na_{pol}\gy\right)
,\quad
j_\gj
=
-j/R
,\eqbl
\bqbl
\lbq{eq:ContinuityEq}
\R{\pa \gr}{\pa t}
&
=
&
-\na\cdot\left(\gr\wsv\right)
+\na\cdot\left(D\na\gr\right)
+S_{D}
,\eqbl
\bqbl
\lbq{eq:VorticityEq}
R\na\cdot\left[R^2\R{\pa}{\pa t}\left(\gr\na_{pol}u\right)\right]
&
=
&
\R{1}{2}\left\{R^2\left|\na_{pol}u\right|^2,R^2\gr\right\}
+\left\{R^4\gr\gw,u\right\}
\nonumber
\\
&&
-R\na\cdot\left[R^2\na_{pol}u\na\cdot\left(\gr\wsv\right)\right]
+\left\{\gy,j\right\}
\nonumber
\\
&
&
-\R{F_0}{R}\R{\pa j}{\pa\gj}
+\left\{P,R^2\right\}
+R\gm_\bot\left(T_e\right)\na_{pol}^2\gw
,\eqbl
\bqbl
\lbq{eq:VorticityDef}
\gw
&
=
&
\R{1}{R}\R{\pa}{\pa R}\left(R\R{\pa u}{\pa R}\right)
+\R{\pa^2 u}{\pa Z^2}
,\eqbl
\bqbl
\lbq{eq:ParallelEq}
B^2\R{\pa}{\pa t}\left(\gr v_\|\right)
&
=
&
-\R{1}{2}\gr\R{F_0}{R^2}\R{\pa}{\pa \gj}\left(v_\|B\right)^2
-\R{\gr}{2R}\left\{B^2v_\|^2,\gy\right\}
-\R{F_0}{R^2}\R{\pa P}{\pa \gj}
\nonumber
\\
&
&
+\R{1}{R}\left\{\gy,P\right\}
-B^2\na\cdot\left(\gr\wsv\right)v_\|
+B^2\gm_\|\na_{pol}^2v_\|
.\eqbl
\bqbl
\lbq{eq:i_TemperatureEq502}
\R{\pa}{\pa t}P_i
&
=
&
-\wsv\cdot \na P_i
-\gg P_i\na\cdot\wsv
+\na\cdot\left(\gk_\bot\na_\bot T_i
+\gk_{i,\|}\na_\| T_i\right)
\nonumber
\\
&&
+\R{\gg-1}{2}\wsv\cdot\wsv S_{D}
+\left(\gg-1\right)\gm_\|\left[\na_{pol}\left(v_\|B\right)\right]^2
+n_i\left(\pa_t T_i\right)_{c,e}
,\eqbl
\bqbl
\lbq{eq:e_TemperatureEq502}
\R{\pa}{\pa t}P_e
&
=
&
-\wsv\cdot \na P_e
-\gg P_e\na\cdot\wsv
+\na\cdot\left(\gk_\bot\na_\bot T_e
+\gk_{e,\|}\na_\| T_e\right)
\nonumber
\\
&&
+\R{\gg-1}{R^2}\gh\left(T_e\right)j^2
-\left(\gg-1\right)P_{ion}
+n_e\left(\pa_t T_e\right)_{c,i}
.\eqbl
Here, the Poisson bracket is defined as $\left\{f,g\right\}\equiv R\left(\na f\times\na g\right)\cdot\na\gj$ and $\na_{pol}$ represents gradient in the poloidal plane. Eq.\,(\rfq{eq:MagField}) and (\rfq{eq:VelField}) are the reduced MHD representation of the magnetic and velocity field. Eq.\,(\rfq{eq:InductionEq}) is the induction equation, with $\gh$ the Spitzer resistivity \cite{SpitzerBook} calculated from the local fluid electron temperature. A more rigorous consideration would include the contribution from the hot-tail electrons. However, for D2 injection the post-SPI fluid electron temperature is usually of $\mathcal{O}\left(100eV\right)$, thus the resistivity effect is subdominant anyway. Eq.\,(\rfq{eq:ContinuityEq}) is the continuity equation and $S_D$ is the ablative density source, we will discuss the details of $S_D$ calculation later. Eq.\,(\rfq{eq:VorticityEq})-(\rfq{eq:ParallelEq}) are the perpendicular and the parallel momentum equations. Lastly, Eq.\,(\rfq{eq:i_TemperatureEq502}) and (\rfq{eq:e_TemperatureEq502}) are the ion and electron pressure equations. Here, $P_{ion}$ is the ionization power from the assimilated deuterium, while $n_i\left(\pa_t T_i\right)_{c,e}$ and $n_e\left(\pa_t T_e\right)_{c,i}$ are the collisional thermalization terms between the fluid ion and electron species. Note that, rigorously, Eq.\,(\rfq{eq:e_TemperatureEq502}) should include the bulk fluid contribution and moments of hot electrons separately. However, since the hot electron density decrease quickly as is shown in Fig.\,\ref{fig:15keVnosink}, Fig.\,\ref{fig:20keV} and Fig.\,\ref{fig:5keV}, and we are not concerned with the fast ideal instabilities in this axisymmetric investigation, we hereby only consider the finite thermalization time effect in the ablation term due to its weak dependence on the hot electron density.
Thus Eq.\,(\rfq{eq:e_TemperatureEq502}) effectively governs the total thermal energy density, and we assume that the energy carried by those hot electrons lost to the self-limiting effect are transferred to the bulk cold electron fluid, so that there is no corresponding sink term in Eq.\,(\rfq{eq:e_TemperatureEq502}).

We represent the hot-tail electron population as a function of the normalized flux $\bgy$. Due to their rapid decrease in density as shown in Section \ref{ss:NumericalThermalization}, we neglect their contribution to the pressure terms, and only consider their impact on the ablation source $S_D$ which has a weak density dependence but a strong dependence on the ambient electron temperature. For deuterium fragments, the mass ablation rate can be written as \cite{Zhang2020NF}:
\bqbl
\lbq{eq:AblrationRate}
G\left[g/s\right]
=
\gl_{D2}\gz_B\left(\R{T_e\left[eV\right]}{2000}\right)^{5/3}
\left(\R{r_p\left[cm\right]}{0.2}\right)^{4/3}n_e\left[10^{14}cm^{-3}\right]^{1/3}
.\eqbl
Here $\gl_{D2}\simeq 39$, and $\gz_B\simeq\left(\R{2}{\max{\left(B[T],2\right)}}\right)^{0.843}$ is a fitted factor to represent the additional shielding effect in high magnetic field. The ambient electron temperature $T_e$ and density $n_e$ could either be chosen as the local fluid ones or as the hot-tail ones $T_{e,hot}(\bgy)$ and $n_{e,hot}(\bgy)$. We call the former choice the ``fluid treatment'' and the latter one ``hot-tail treatment''. Note that when using the ``hot-tail treatment'' and we find that the ablation rate from the hot-tail electrons is lower than that from the fluid electrons, we deem the hot-tails completely thermalized and use the fluid electron result instead.
The ablation source $S_D$ is then calculated by $G$ and distributed around the fragment with a poloidal gaussian shape in our axisymmetric simulations:
\bqbl
S_D
\propto
\exp{\left(-\R{\left(R-R_{f}\right)^2+\left(Z-Z_{f}\right)^2}{\gD r^{2}_{NG}}\right)}
.\eqbl
Here, we choose the deposition radius radius to be $\gD r_{NG}=4cm$. We do not find significant difference in the numerical results by varying this parameter in the range of $2cm$ to $8cm$.

The only question remaining is how should the hot-tail electron temperature and density evolution be modelled.
According to our conclusion in Section \ref{ss:AnalyticalThreshold}, with the self-limiting effect, we would keep $T_{e,hot}(\bgy)$, thus $\gq(\bgy)$, constant while evolve $n_{e,hot}(\bgy)$ according to the two mechanisms discussed in Section \ref{ss:AnalyticalThreshold}:
\bqbl
\lbq{eq:JOREKSink}
\R{\pa}{\pa t}n_{e,hot}(\bgy)
&
=
&
-\R{1}{2\sqrt{2}}\R{e^4n_{e,cold}\ln{\gL}}{\gp\ge_0^2m_e^2v_0^3}
\R{\gq_0^{-3/2}}{2S^{1/3}+e^{S^{2/3}}\sqrt{\gp}erfc(S^{1/3})}n_{e,hot}(t)
\nonumber
\\
&&
-\R{\sum_{\bgy}{\gp r_p^2c}}{\gD V(\bgy)}\R{\gq\left(1+\gq\right)}{K_2\left(1/\gq\right)}e^{-1/\gq}n_{e,hot}(\bgy)
.\eqbl
To facilitate our numerical treatment, we treat the flux surfaces as discrete bundles of flux surfaces with small but finite width.
Here, normalized time $S(\bgy)=0$ corresponds to when the fragments first arrive on a bundle of flux surfaces around normalized flux $\bgy$. Further, $\sum_{\bgy}\cdots$ represents summing over all fragments on that bundle of flux surfaces, and $\gD V(\bgy)$ is the volume enclosed by those flux surfaces.
On the other hand, it is also of interest to study the hot-tail impact without the self-limiting effect and compare the result. In this limit only the first term on the right hand side of Eq.\,(\rfq{eq:JOREKSink}) is kept and $\gq(t)$ follows a fitted curve according to the Smith result \cite{Smith2008POP}.

Eq.\,(\rfq{eq:MagField})-(\rfq{eq:e_TemperatureEq502}), combined with the ``fluid'' or ``hot-tail'' ablation source described above, then forms a complete set of governing equation which we will use to investigate the hot tail impact with and without the self-limiting effect.

\subsection{Target equilibrium and injection configuration}
\label{ss:Equilibrium}

\begin{figure*}
\centering
\noindent
\btbl{c}
\parbox{6.5in}{
    \includegraphics[scale=0.45]{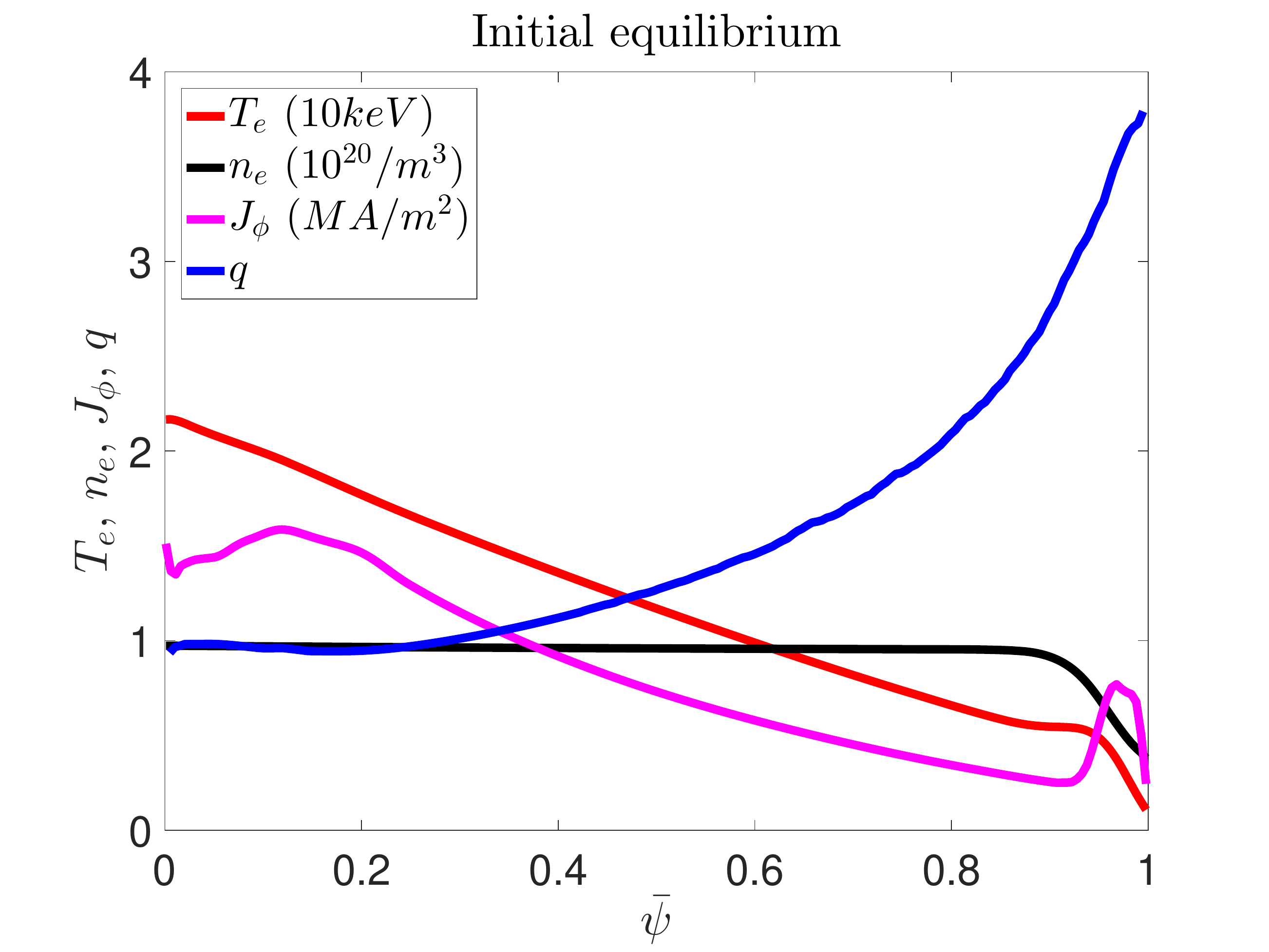}
}
\etbl
\caption{The initial equilibrium. $\gY_N$ is the normalized poloidal flux.}
\label{fig:InitEq}
\end{figure*}

We consider an ITER H-mode scenario as the target plasma \cite{Kim2018NF}. This plasma consists of half deuterium and half tritium, however for the purpose of this study it suffices to assume the plasma is of pure deuterium. The toroidal magnetic field is $B_T=5,3T$, the plasma current is $I_p=15MA$. The initial equilibrium is shown in Fig.\,\ref{fig:InitEq}, where the initial electron density, temperature, as well as the current density and the safety factor is given as functions of the normalized flux $\bgy$.  We assume the the perpendicular particle diffusivity $D_\perp\simeq7m^2/s$ and the perpendicular heat diffusivity $\gk_\perp\simeq 10m^2/s$. The Braginskii heat conductivity is used for the parallel heat conduction \cite{Braginskii1965RPP}, but with a upper ceiling corresponding to the free streaming heat conduction in the collisionless limit.

We only consider the D2 SPI case here. The fragments are injected from the outer mid-plane with vertex angle $20$ degree, and the fragments are injected from outer mid-plane inward along the major radial direction. To demonstrate the hot-tail effect for a variety of injection configurations, five different sets of SPI parameters are chosen, as is detailed in Table \ref{tab:SPIconfig}, including the name of the configuration, the injection amount, the fragment number and the injection velocity. The fragment radius distribution satisfy the Statistical Fragmentation model described by Eq.\,(\rfq{eq:ParkP}), while we used a flat velocity distribution function. We assume all injected materials are converted into fragments. Among these configurations, the Reference case, the Finer Fragment case and the Slow case is close to the deuterium amount a single ITER DMS injector can provide, and their velocities correspond to the higher and lower bound of injection velocity for ITER SPIs, so that we call them ``ITER-like''.

\begin{table*}
\centering
\noindent
\btbl{|c|c|c|c|}
\hline
Notation & Injected D Atoms & Fragment Number & Velocity\\
\hline
Reference case & $2.1\times10^{24}$ & $1000$ & $500\pm 200m/s$ \\
\hline
Finer Fragment case & $1.05\times10^{24}$ $(\times 2)$ & $1000$ $(\times 2)$ & $500\pm 200m/s$ \\
\hline
Small \& Compact case & $10^{24}$ & $1000$ & $500\pm 100m/s$ \\
\hline
Slow case & $2.1\times10^{24}$ & $1000$ & $150\pm 50m/s$ \\
\hline
Unshattered Pellet case & $10^{24}$ & $1$ & $500m/s$ \\
\hline
\etbl
\caption{The injection parameters for the SPI considered in this study. }
\label{tab:SPIconfig}
\end{table*}

\subsection{JOREK simulation of the injection assimilation and penetration with and without the finite thermalization effect}
\label{ss:JOREKResult}

\begin{figure*}
\centering
\noindent
\btbl{c}
\parbox{5.0in}{
    \includegraphics[scale=0.45]{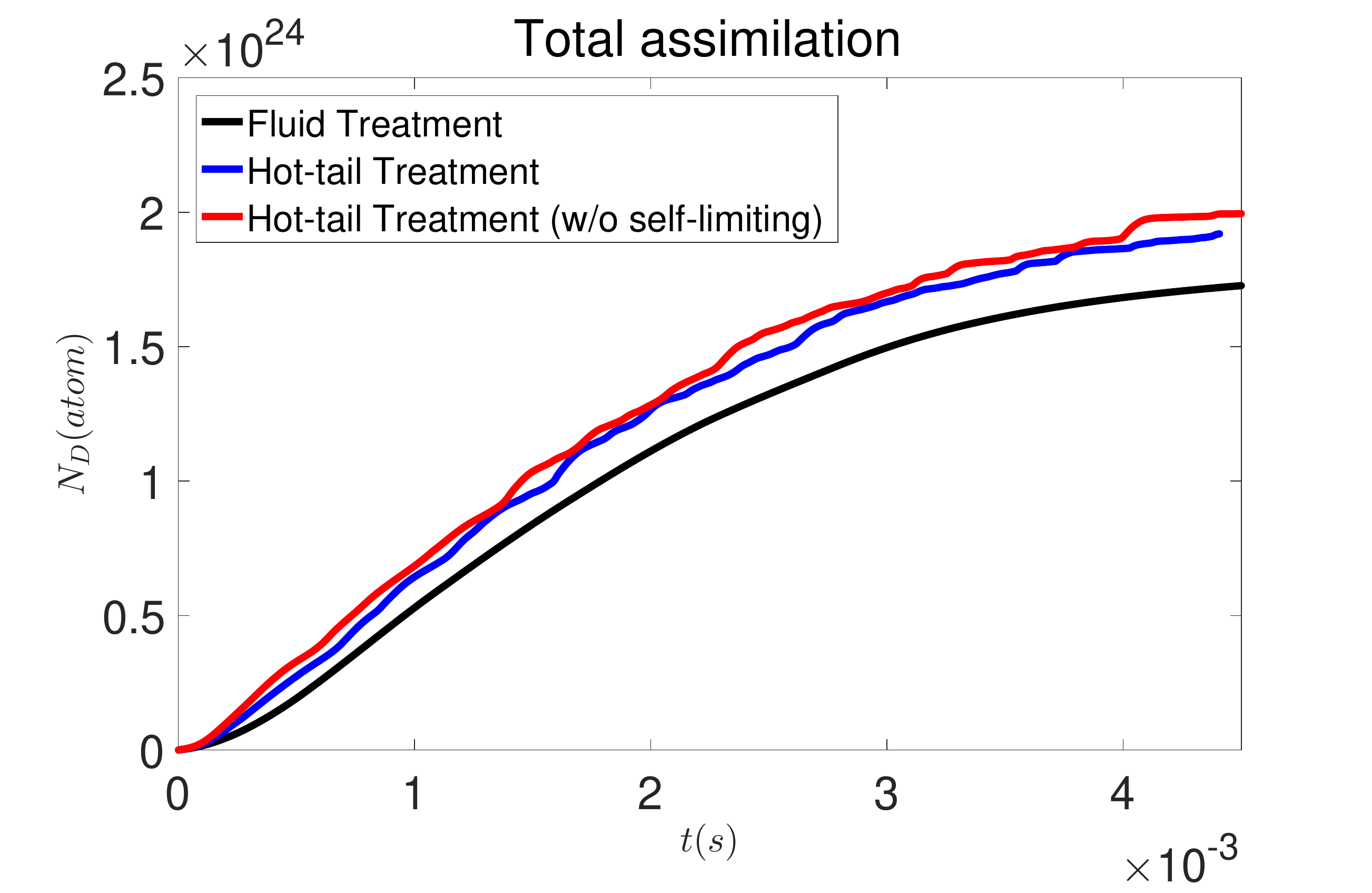}
}
\etbl
\caption{The total assimilation comparison for the Reference Case between the fluid treatment (black), the hot-tail treatment (blue) and the hot-tail treatment without the self-limiting effect (red).}
\label{fig:15keVAblCompare}
\end{figure*}

We first look at the Reference case. The total deuterium assimilation as a function of time is shown in Fig.\,\ref{fig:15keVAblCompare}. The assimilation for the fluid treatment is shown in black, while that for the hot-tail treatment, with and without the self-limiting effect, are shown in blue and red respectively. It is apparent that the hot-tail indeed result in enhanced ablation, while the self-limiting effect only slightly mitigate such enhancement for this configuration. Most of the fragments are ``burnt up'' before they reach the magnetic axis at $t\simeq 4ms$ for the hot-tail treatment, while some fragments survived as is suggested by the black solid line reaching plateau while some $3\times10^{23}$ atoms are still in the solid fragments. However, the overall assimilation enhancement by hot-tail effect is of the next order compared with the simple fluid treatment.

\begin{figure*}
\centering
\noindent
\btbl{cc}
\parbox{2.5in}{
    \includegraphics[scale=0.275]{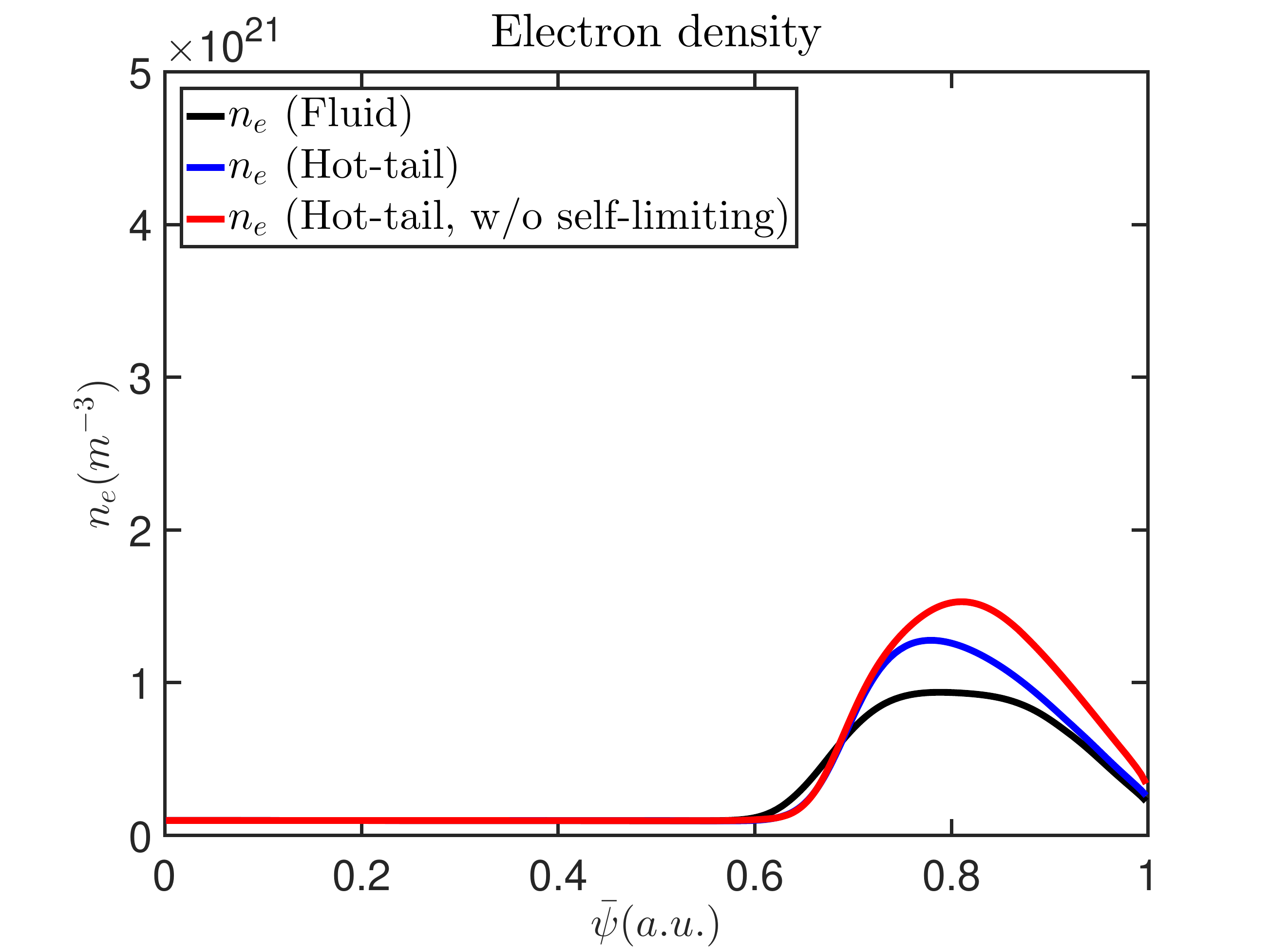}
}
&
\parbox{2.5in}{
	\includegraphics[scale=0.275]{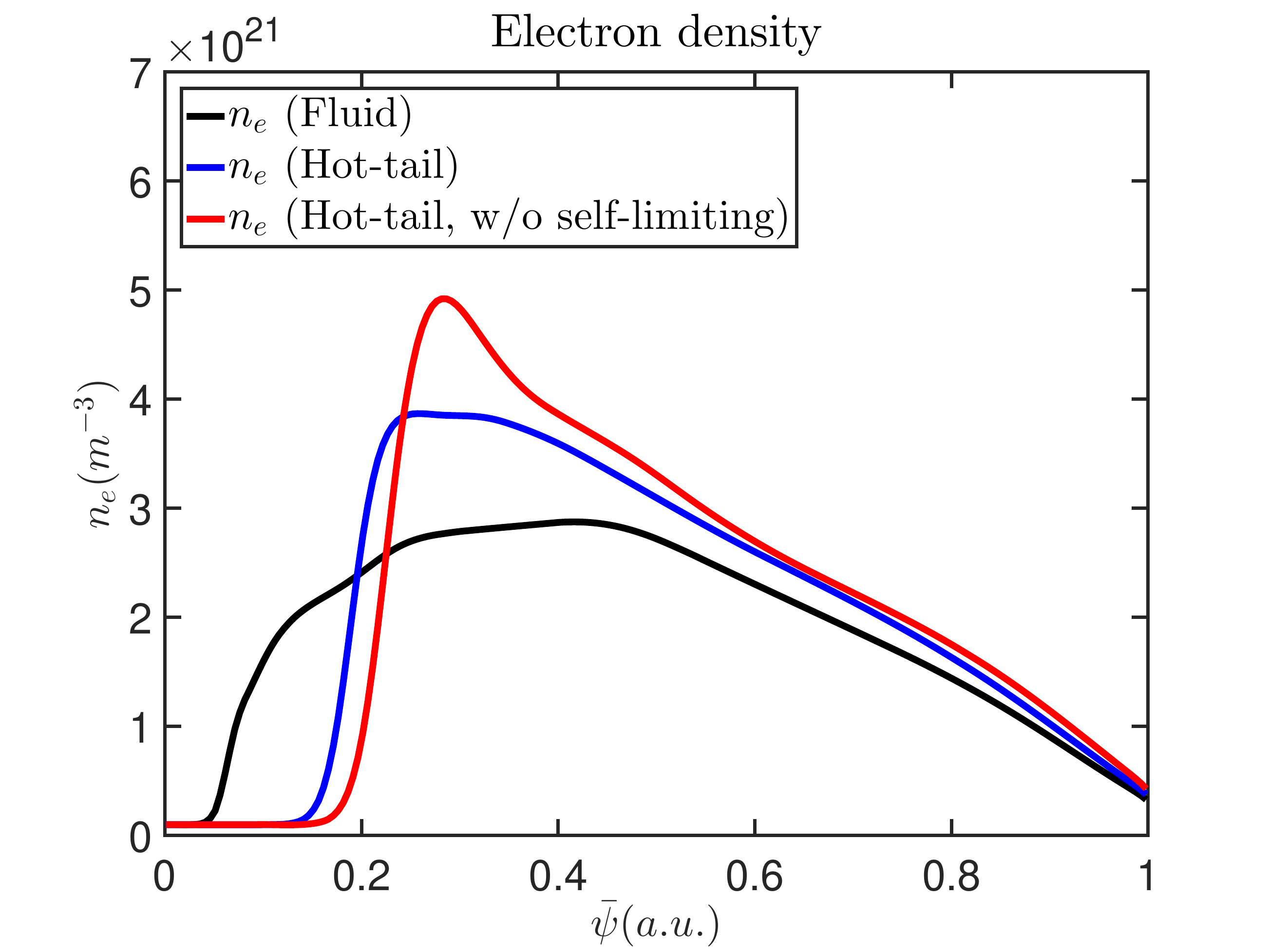}
}
\\
(a)&(b)
\etbl
\caption{The electron density profile at (a) $t=0.64ms$ and (b) $t=2.56ms$. The black lines represent the fluid treatment, the blue lines represent the hot-tail treatment and the red lines represent the hot-tail treatment without the self-limiting effect.}
\label{fig:ne_compare_reference_case}
\end{figure*}

Apart from the total assimilation, we are also concerned with the hot-tail impact on the injection penetration, as the material deposition and cooling on major rational surfaces, such as the $q=1$ or $q=2$ surface, could have significant impact on the MHD activity as well as the convective transport and core mixing of the injected materials \cite{Di2018NF,Di2021NF}. The electron density profiles of the Reference Case at $t=0.64ms$ and $t=2.56ms$ after the injection are shown in Fig.\,\ref{fig:ne_compare_reference_case}. The black lines indicate the fluid treatment result, while the blue and red lines represent the hot-tail treatment with and without the self-limiting effect respectively. For $t=0.64ms$, the vanguard fragments from all three treatments approximately arrive on the $q=2$ surface at $\bgy\simeq0.8$. There is no obvious deviation in the injection penetration between the three cases, as most of the fragments still survive and the ablation enhancement mostly manifest itself in the difference in the density peak. Although a closer inspection reveals that there are indeed small differences in the density peak position. At time $t=2.56ms$, the vanguard fragments from the fluid treatment have already crossed the $q=1$ surface at $\bgy\simeq0.2$ and deposited some materials within, while that from the hot-tail treatments have just barely reaches the $q=1$ surface. The deviation in injection penetration is already significant by this time, as more vanguard fragments are ``burnt up'' due to the enhanced ablation in the hot-tail treatments. The self-limiting effect provides some slight mitigation of the hot-tail ablation enhancement, as indicated by the slightly deeper penetration of the blue solid curve. We will see later that such mitigation is subject to the injection configuration, and with denser fragment plume the mitigation becomes more significant.

\begin{figure*}
\centering
\noindent
\btbl{c}
\parbox{5.0in}{
    \includegraphics[scale=0.45]{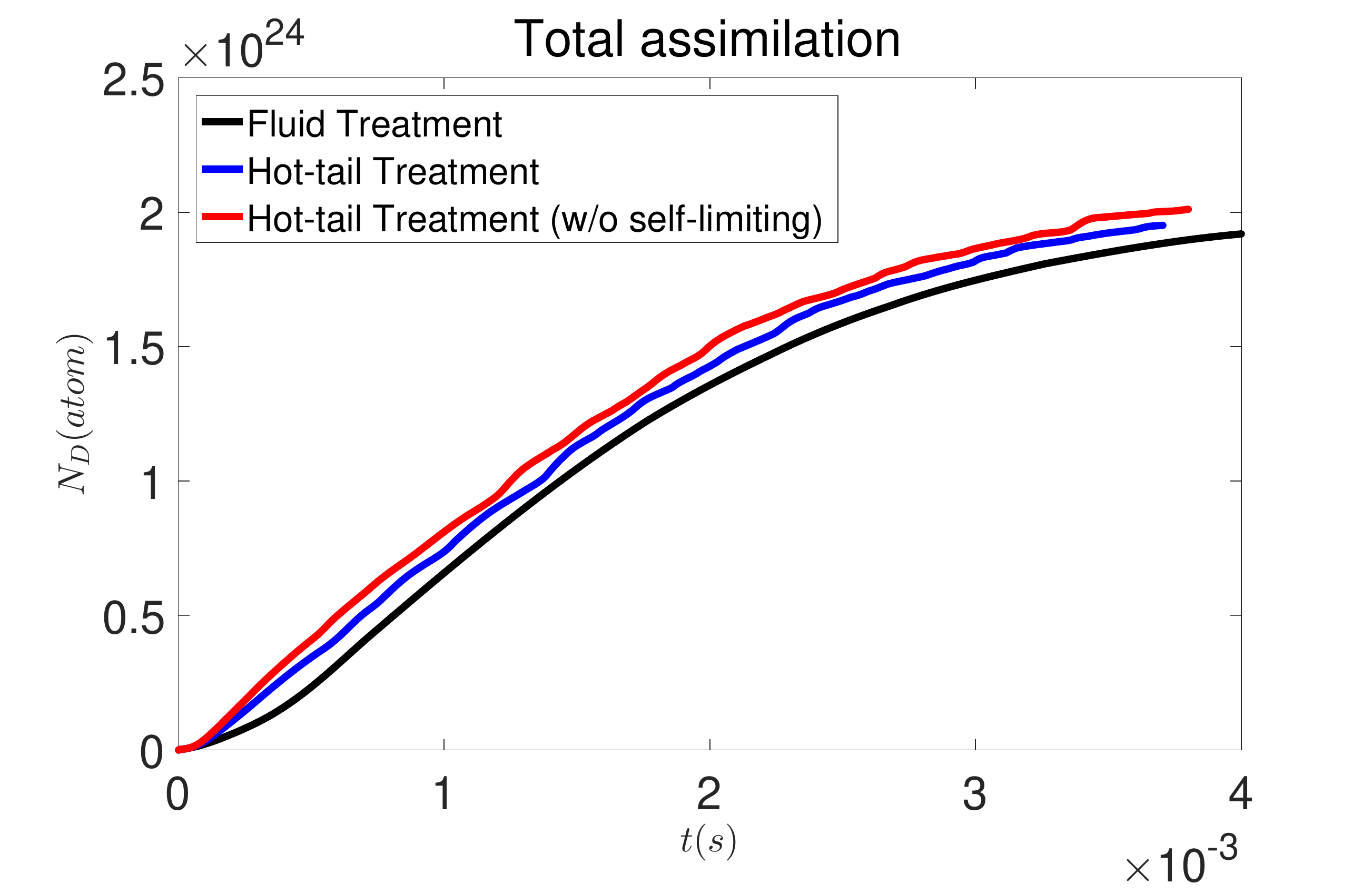}
}
\etbl
\caption{The total assimilation comparison for the Finer Fragment case between the fluid treatment (black), the hot-tail treatment (blue) and the hot-tail treatment without the self-limiting effect (red).}
\label{fig:DoubleAblCompare}
\end{figure*}

As an example of the hot-tail effect for different injection configuration, we look at the fluid and hot-tail treatment comparison for the Finer Fragment case in Table \ref{tab:SPIconfig}. The total injection amount and injection velocity is the same with the Reference case, but here we use two injectors so that the injection amount from each injector is half of that in the Reference case. The fragment size follows the Statistical Fragmentation model for each injector respectively. This effectively result in finer fragments, thus denser fragment plume on each flux surface they pass by. The total assimilation history is shown in Fig.\,\ref{fig:DoubleAblCompare}. Again some assimilation enhancement is seen comparing the red, blue and black lines, although in this case the difference is even less pronounced compared with that in Fig.\,\ref{fig:15keVAblCompare}.

\begin{figure*}
\centering
\noindent
\btbl{cc}
\parbox{2.5in}{
    \includegraphics[scale=0.275]{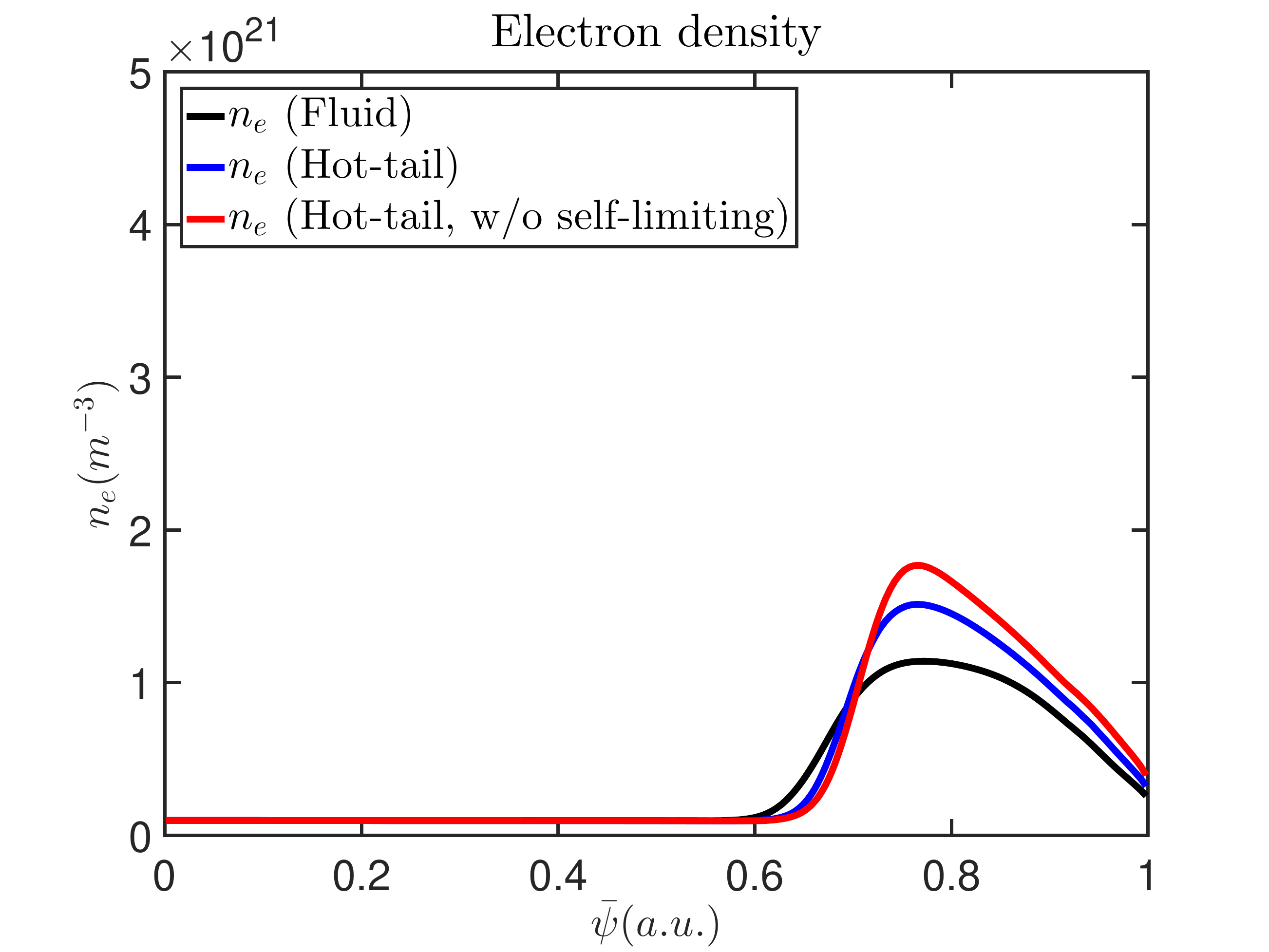}
}
&
\parbox{2.5in}{
	\includegraphics[scale=0.275]{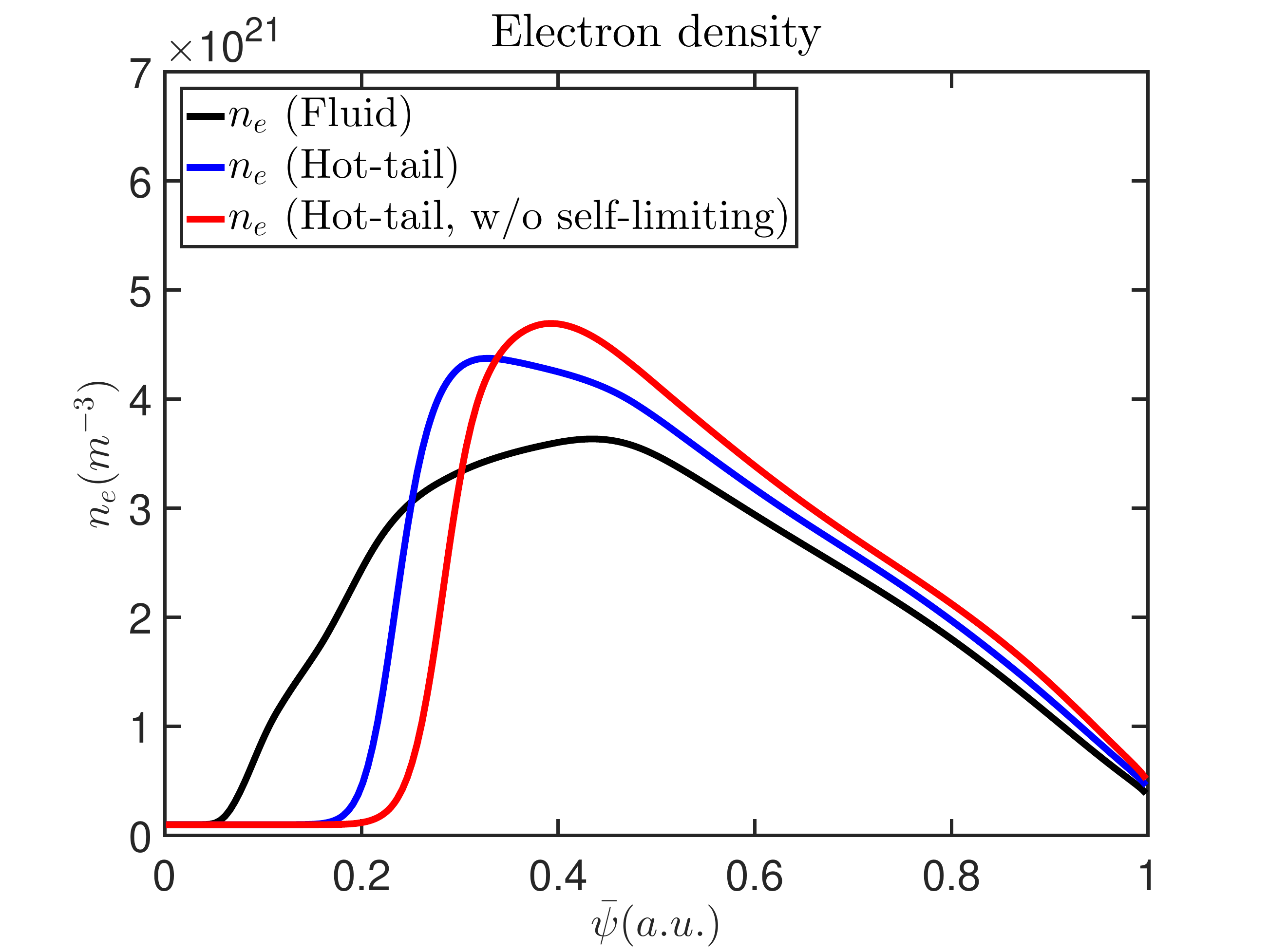}
}
\\
(a)&(b)
\etbl
\caption{The electron density profile at (a) $t=0.64ms$ and (b) $t=2.56ms$ for the Finer Fragment case. The black lines represent the fluid treatment, the blue lines represent the hot-tail treatment and the red lines represent the hot-tail treatment without the self-limiting effect.}
\label{fig:ne_compare_double_injection}
\end{figure*}

The electron density profiles shown in Fig.\,\ref{fig:ne_compare_double_injection} also exhibit similar trend with that of the Reference case. At $t=0.64ms$, the electron density rise on the $q=2$ surface at $\bgy\simeq 0.8$ is shown in Fig.\,\ref{fig:ne_compare_double_injection}(a). As in the Reference case, the hot-tail effect manifest as a increase density peak at this early stage, while the difference between the hot-tail treatments and the fluid treatment in the penetration is insignificant. Later on when the fragments reach $q=1$ surface at $\bgy\simeq 0.2$ as shown in Fig.\,\ref{fig:ne_compare_double_injection}(b), the difference in the penetration becomes more apparent. Like in the previous case, both hot-tail treatments shows a steeper density gradient toward the plasma axis, while the fluid treatment density gradient is more relaxed. This is due to more small radius vanguard fragments survived in the fluid treatment thus penetrating deeper. The self-limiting effect is shown to mitigate such deviation in penetration, although not completely. Comparing Fig.\,\ref{fig:ne_compare_double_injection}(b) and Fig.\,\ref{fig:ne_compare_reference_case}(b), such mitigation is stronger in the Finer Fragment case due to the denser fragment plume. The density profile in the wake of the density peak shows more or less similar behavior.

\begin{figure*}
\centering
\noindent
\btbl{c}
\parbox{5.0in}{
    \includegraphics[scale=0.45]{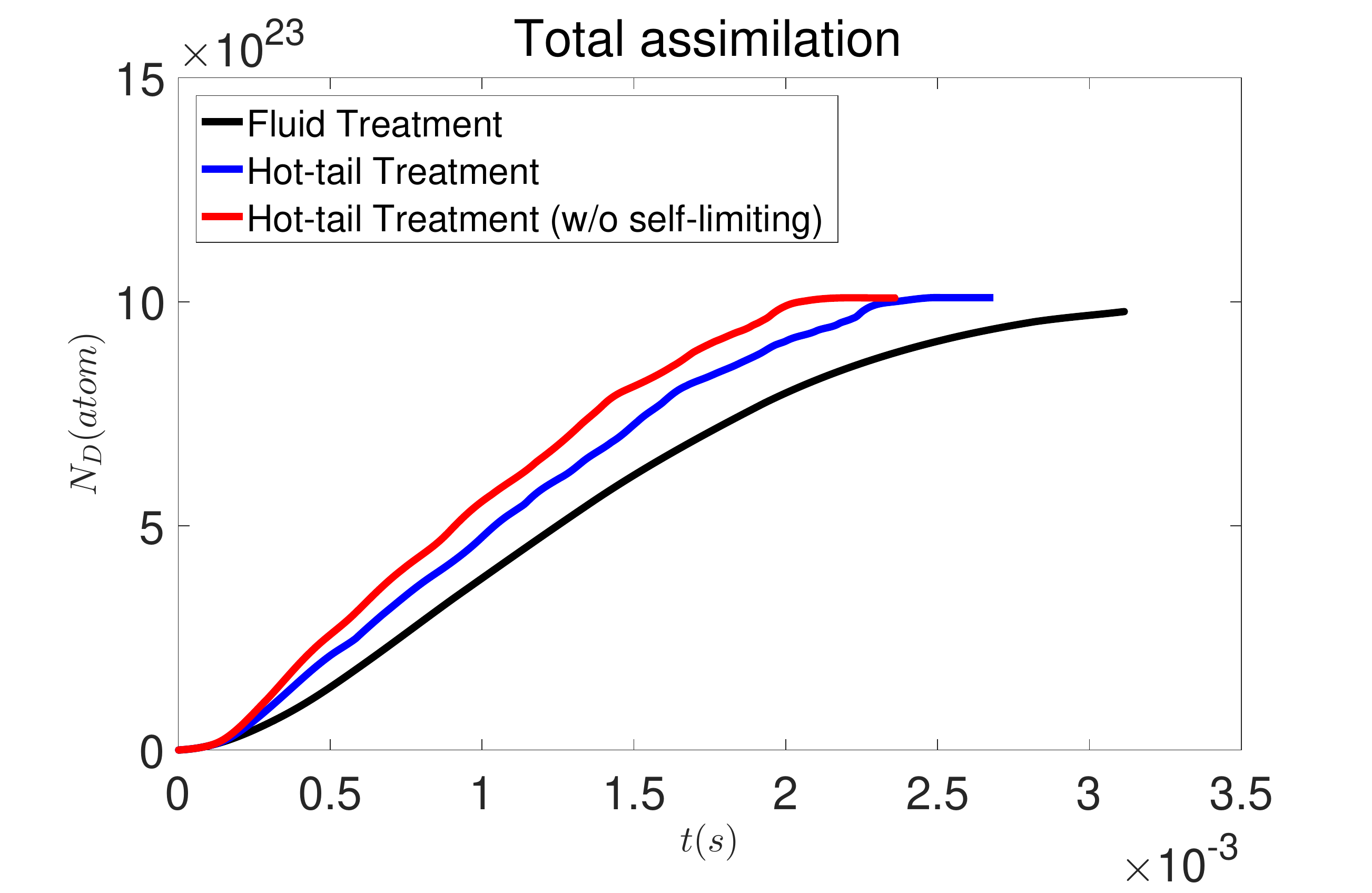}
}
\etbl
\caption{The total assimilation comparison for the Small \& Compact case between the fluid treatment (black), the hot-tail treatment (blue) and the hot-tail treatment without the self-limiting effect (red).}
\label{fig:SCAblCompare}
\end{figure*}

Another example is the Small \& Compact case as detailed in Table \ref{tab:SPIconfig}. The total assimilation as functions of time are shown in Fig.\,\ref{fig:SCAblCompare}. Without the self-limiting effect, the hot-tail impact is more significant for the Small \& Compact case as now one fourth of the solid mass still remain in the black curve while all is ``burnt up'' in the red one. This enhancement of the hot-tail impact is likely due to the compactness of the fragment plume, so that the fragments following the vanguards spend shorter time to reach the flux surfaces cooled down by the vanguards, and the collisional thermalization is less complete when they arrive compared to the Reference Case, hence the stronger additional ablation.
Meanwhile, the self-limiting mitigation is also more potent as can be seen by the more obvious separation between the blue and red curve in Fig.\,\ref{fig:SCAblCompare}, as opposed to the case in Fig.\,\ref{fig:15keVAblCompare}. This mitigation is also due to the compactness of the fragments, which increase the $\sum_{\bgy}{\gp r_p^2}$ term in Eq.\,(\rfq{eq:JOREKSink}). In this sense, longitudinal spreading of the fragment plume during SPI shows self-regulating behavior on the hot-tail effect.

\begin{figure*}
\centering
\noindent
\btbl{cc}
\parbox{2.5in}{
    \includegraphics[scale=0.275]{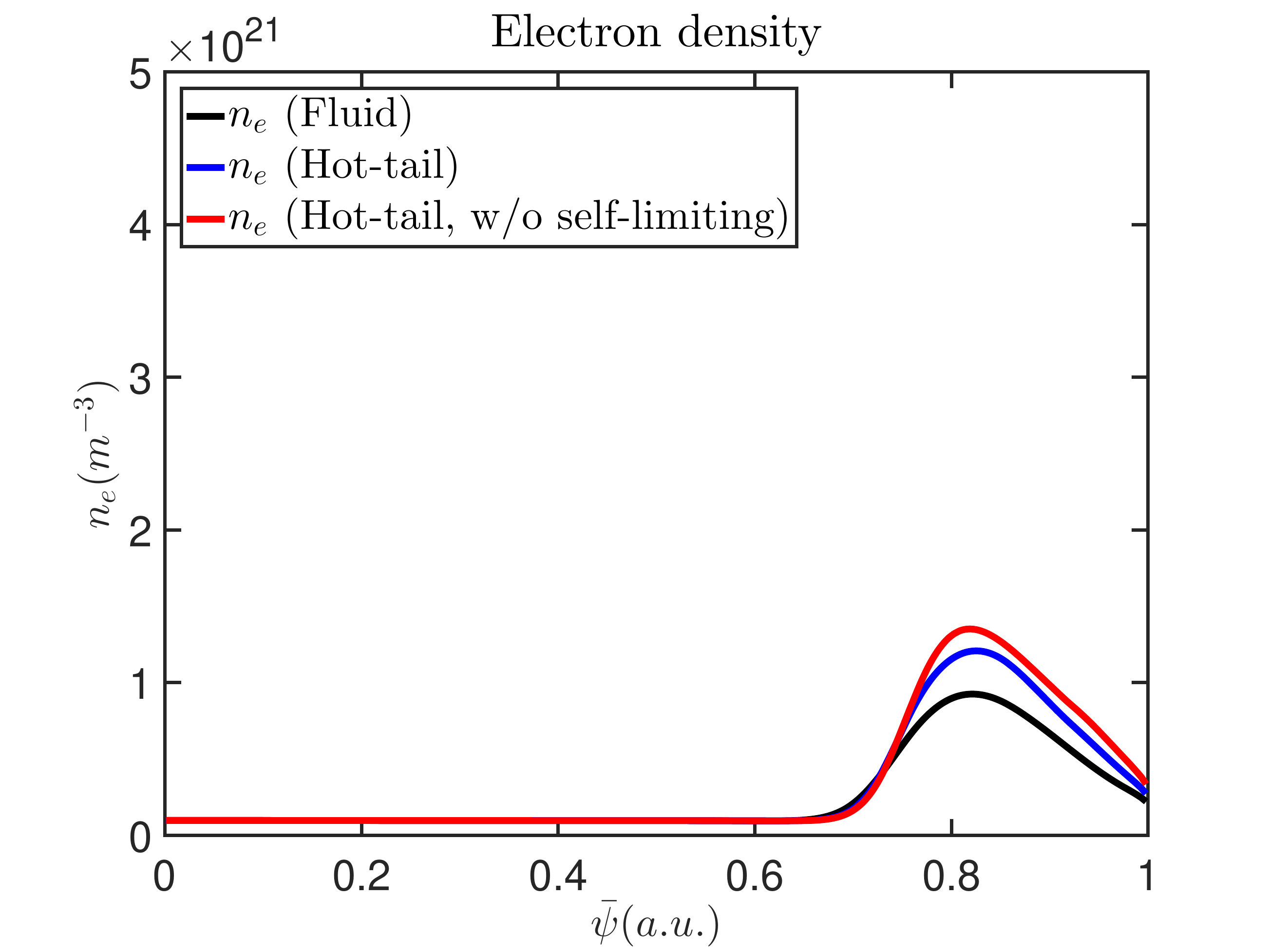}
}
&
\parbox{2.5in}{
	\includegraphics[scale=0.275]{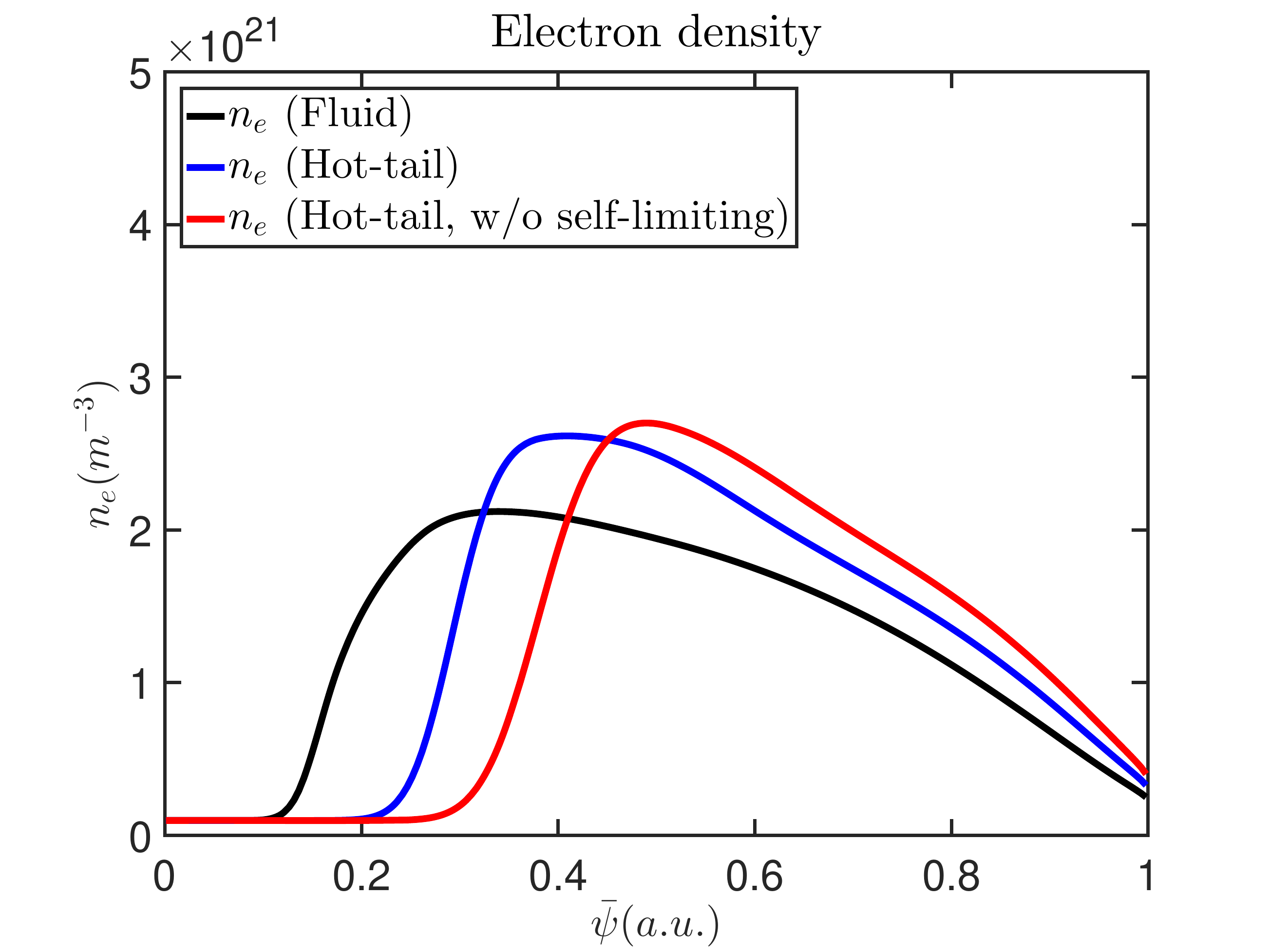}
}
\\
(a)&(b)
\etbl
\caption{The electron density profile at (a) $t=0.64ms$ and (b) $t=2.46ms$. The black lines represent the fluid treatment, the blue lines represent the hot-tail treatment and the red lines represent the hot-tail treatment without the self-limiting effect.}
\label{fig:ne_compare_SC_case}
\end{figure*}

The density profile comparison for the Small \& Compact case is shown in Fig.\,\ref{fig:ne_compare_SC_case}. At $t=0.64ms$, Fig.\,\ref{fig:ne_compare_SC_case}(a) looks almost the same with Fig.\,\ref{fig:ne_compare_reference_case}(a) although the density peak in the former is somewhat lower due to the smaller injection quantity thus the smaller fragment radius. Further at $t=2.46ms$ shown in Fig.\,\ref{fig:ne_compare_SC_case}(b), the fragments from the hot-tail treatments all ``burnt up'' just before they reach the $q=1$ surface at $\bgy\simeq0.2$, while some small amount of atoms still remain solid for the fluid treatment. Once again, the hot-tail treatments show shallower penetration compared with that of the simple fluid result, especially if the self-limiting effect is absent. With the self-limiting effect, almost half of the above hot-tail effect is mitigated.

\begin{figure*}
\centering
\noindent
\btbl{c}
\parbox{5.0in}{
    \includegraphics[scale=0.45]{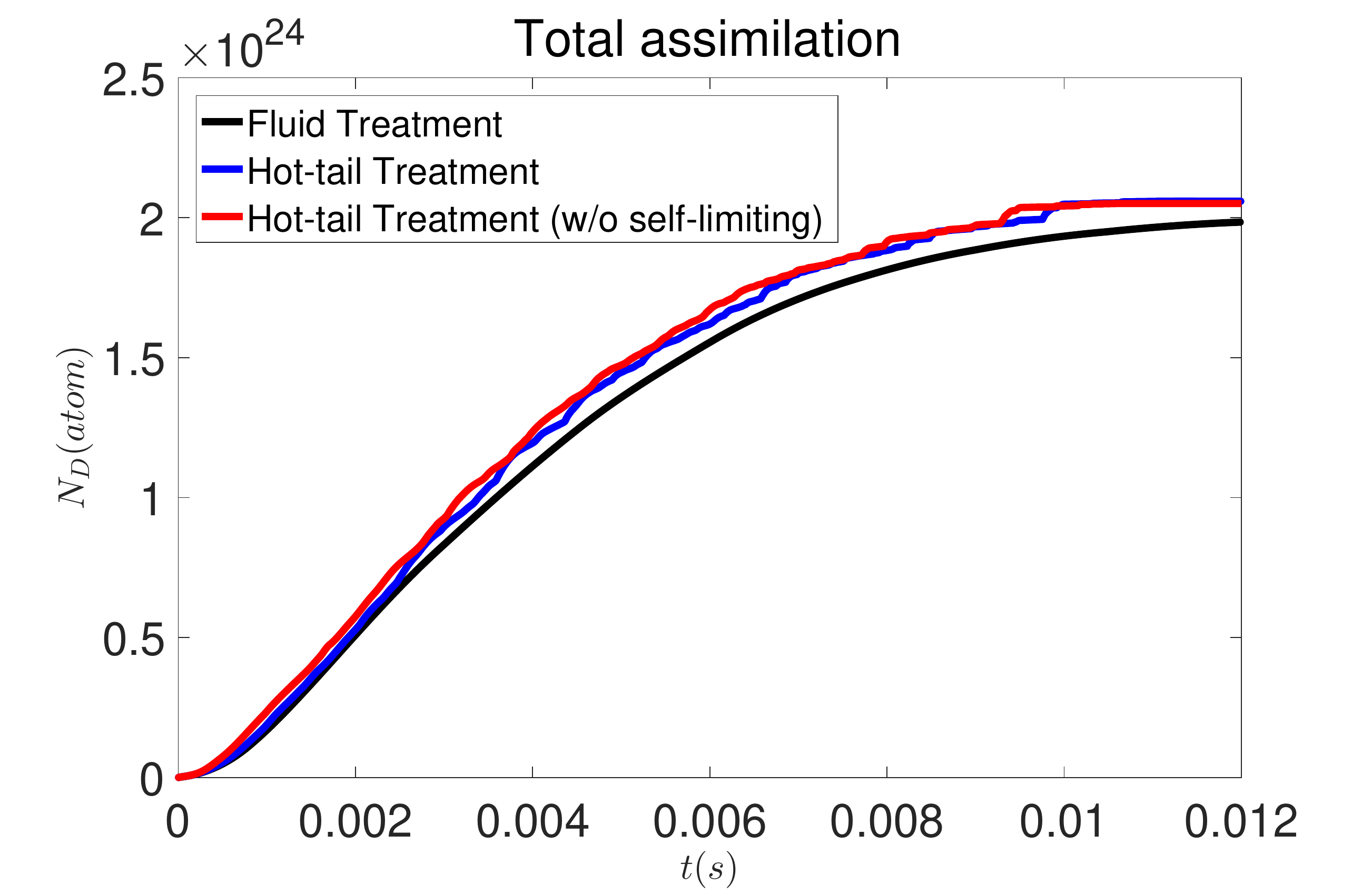}
}
\etbl
\caption{The total assimilation comparison for the Slow case between the fluid treatment (black) and the hot-tail treatment (blue) and the hot-tail treatment without the self-limiting effect (red).}
\label{fig:SlowAblCompare}
\end{figure*}

The injection velocity also plays a strong role in the strength of the hot-tail effect. The comparison between the fluid and hot-tail treatments for the Slow case is shown in Fig.\,\ref{fig:SlowAblCompare}. In contrast to cases shown in Fig.\,\ref{fig:15keVAblCompare} and Fig.\,\ref{fig:SCAblCompare}, the hot-tail effect only have very minor impact to the total assimilation. The case with and without the self-limiting effect are almost indistinguishable. This weakness of the hot-tail effect is due to the slow injection velocity causing the following fragments taking longer time to reach flux surfaces cooled down by the vanguards, so that the thermalization is more complete as they arrive. Hence the hot-tail effect on assimilation is negligible and the simply fluid treatment could be safely used in this injection configuration. This conclusion is also subject to the plasma geometry as it also contribute to the fragment flying time.

\begin{figure*}
\centering
\noindent
\btbl{cc}
\parbox{2.5in}{
    \includegraphics[scale=0.275]{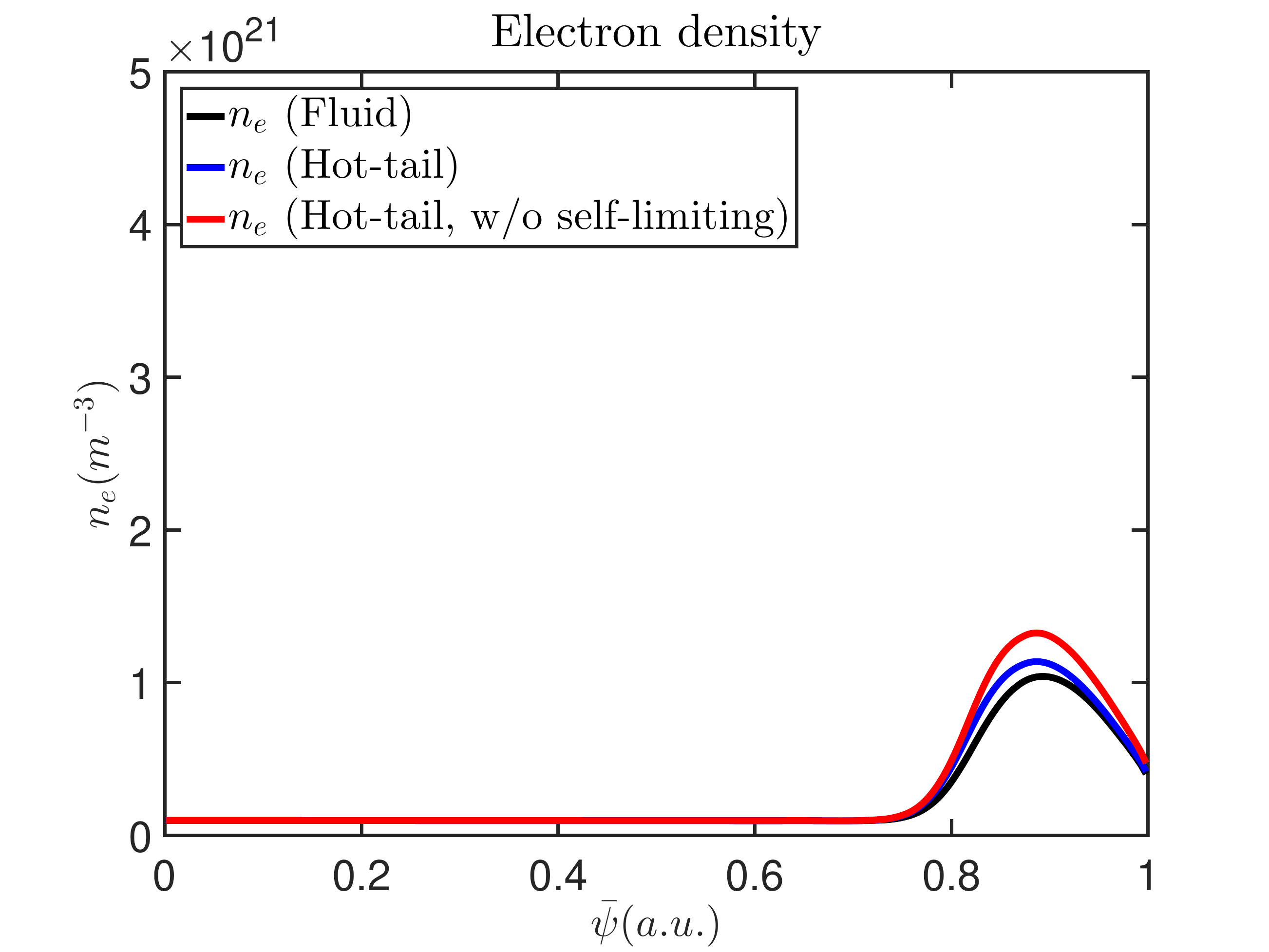}
}
&
\parbox{2.5in}{
	\includegraphics[scale=0.275]{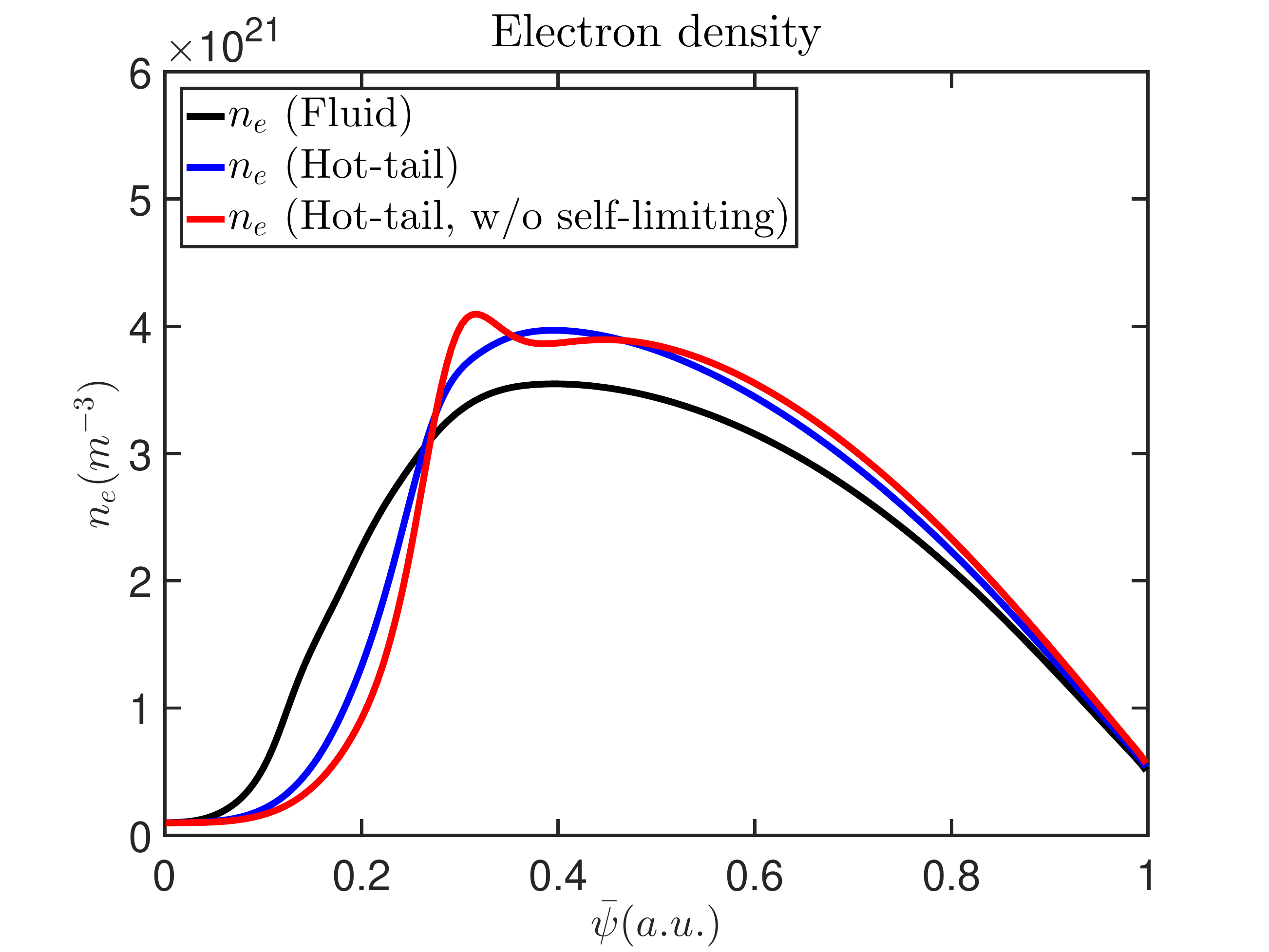}
}
\\
(a)&(b)
\etbl
\caption{The electron density profile at (a) $t=1.25ms$ and (b) $t=9.59ms$. The black lines represent the fluid treatment, the blue lines represent the hot-tail treatment and the red lines represent the hot-tail treatment without the self-limiting effect.}
\label{fig:ne_compare_Slow_case}
\end{figure*}

Likewise, the hot-tail effect on the injection penetration is also greatly decreased in the Slow case, as is shown in Fig.\,\ref{fig:ne_compare_Slow_case}. In the early phase as shown in Fig.\,\ref{fig:ne_compare_Slow_case}(a), the density profile is similar between the three treatments, as is the same with the previous cases. Different from the previous cases, however, the separation between the injection penetration of different treatments is still not pronounced late into the injection, as is shown in Fig.\,\ref{fig:ne_compare_Slow_case}(b). The qualitative trend of the fluid treatment being the most penetrative, followed by normal hot-tail treatment and no-sink hot-tail treatment. The difference is negligible however for the Slow case. This behavior is in consistence with the behavior of the total assimilation, and it is safe to say that the hot-tail effect plays little role in this injection configuration.

\begin{figure*}
\centering
\noindent
\btbl{c}
\parbox{5.0in}{
    \includegraphics[scale=0.45]{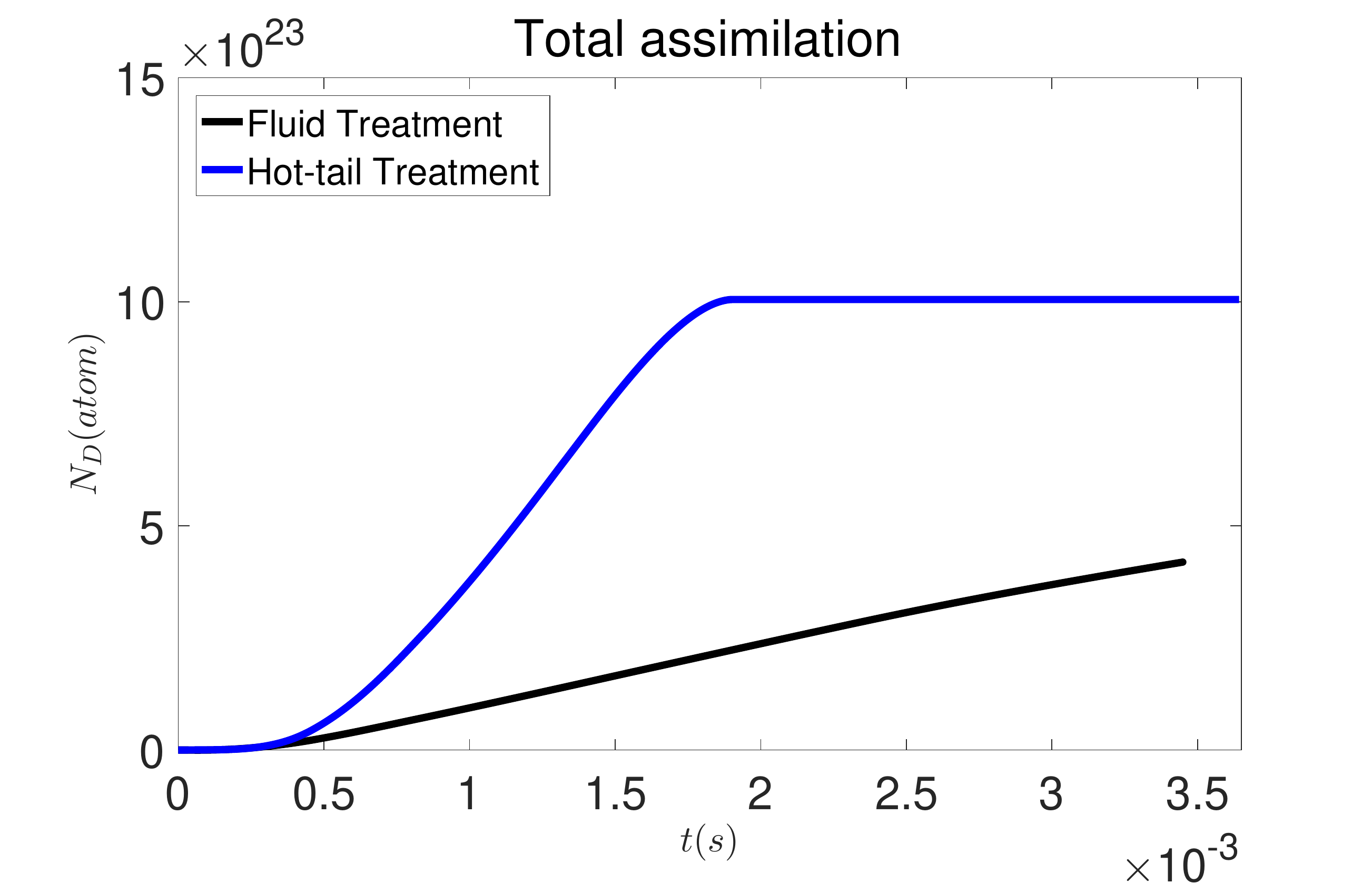}
}
\etbl
\caption{The total assimilation comparison for the Unshattered Pellet case between the fluid treatment (black) and the hot-tail treatment (blue).}
\label{fig:SPAblCompare}
\end{figure*}

Last, we look at an extreme case to demonstrate the maximum possible hot-tail effect on the injection assimilation and penetration. The assimilation comparison for the Unshattered Pellet case as described in Table \ref{tab:SPIconfig} is shown in Fig.\,\ref{fig:SPAblCompare}. The black line is the fluid result while the blue line is the hot-tail with self-limiting. It can be seen that even with self-limiting effect the hot-tail ablation enhancement is still drastic, and the whole pellet ``burnt up'' in the hot-tail treatment while for the fluid treatment merely one fourth of the pellet is assimilated. Such a strong hot-tail effect is likely due to the fact that the Unshattered Pellet always sees the very hot ambient plasma without much time for thermalization. Meanwhile, the small surface-to-volume ratio of the Unshattered Pellet greatly reduces the self-limiting effect. Hence the drastic hot-tail contribution for this extreme case.

\begin{figure*}
\centering
\noindent
\btbl{c}
\parbox{5.0in}{
    \includegraphics[scale=0.45]{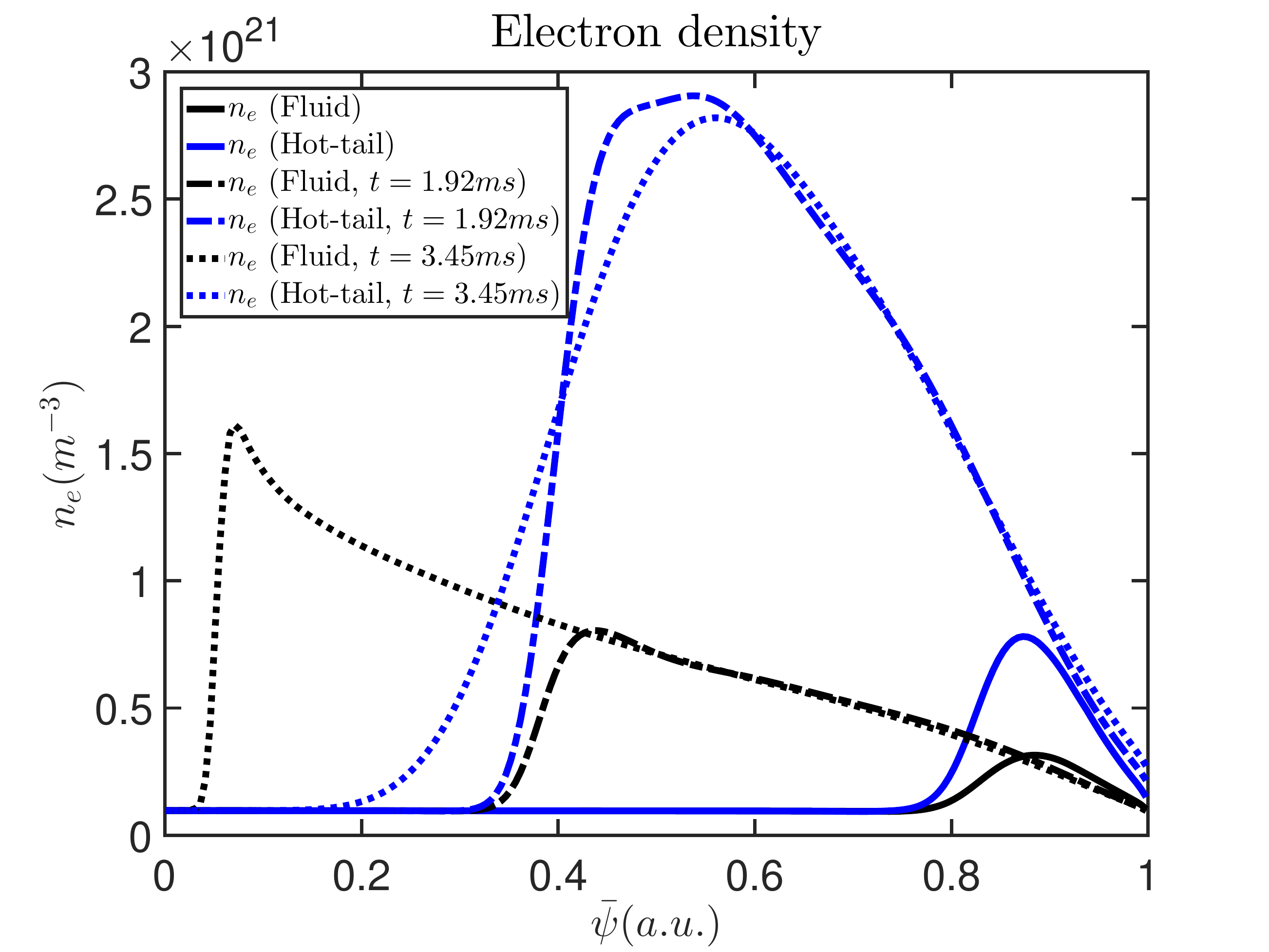}
}
\etbl
\caption{The electron density profiles at different time after the injection. The black lines represent the fluid results while the blue lines represent the hot-tail results.}
\label{fig:ne_compare_SP_case}
\end{figure*}

The comparison of injection penetration at different times for the Unshattered Pellet case are shown in Fig.\,\ref{fig:ne_compare_SP_case}. The black lines are the fluid result and the blue lines are the hot-tail results. The solid lines correspond to $t=0.64ms$, the chained lines correspond to $t=1.92ms$ and the dotted ones correspond to $t=3.45ms$. It can be seen the $t=0.64ms$ profile shares similar shape compared with the result shown in Fig.\,\ref{fig:ne_compare_reference_case}(a) and Fig.\,\ref{fig:ne_compare_SC_case}(a), albeit showing a lower density peak due to being Unshattered Pellet. At $t=1.92ms$, however, the hot-tail result show significantly higher density peak compared with the fluid result, highlighting the drastic strength of the hot-tail effect in the Unshattered Pellet case. This time also corresponds to when the pellet in the hot-tail treatment completely ``burnt up''. Further at $t=3.45ms$, the pellet keeps penetrate deep into the plasma in the fluid treatment, while in the hot-tail treatment only some diffusion occurs since the pellet is already ``burnt up''.

In the above comparisons, we have shown that the finite-time thermalization of the hot-tail electrons could indeed result in enhanced assimilation and shallower fragment penetration for SPIs, while the self-limiting effect tends to mitigate such behavior. The injection velocity is found to have an important role in the strength of the hot-tail effect, with slower injection shows weaker hot-tail effect. While denser fragment plume is found to increase the self-limiting mitigation, the compactness of the plume shows some self-regulated behavior in the overall penetration and assimilation due to the balancing contribution from the finite-time collisional thermalization and the self-limiting effect. Last, an extreme Unshattered Pellet case is used to demonstrate the maximum strength of the hot-tail effect. The pellet is shown to be totally assimilated if the hot-tail effect is considered, while in a simplified fluid treatment most of the pellet mass could survive till it reaches the magnetic axis.

\section{Conclusion and discussion}
\label{s:Conclusion}

In this study, we have investigated the thermalization process of the hot electrons during SPI, and demonstrated the hot-tail electron effect on the SPI assimilation and penetration. Through kinetic simulations of the hot electron distribution function including both the collisional effect and the self-limiting sink, we found the hot electron thermalization exhibits an almost self-similar decay of the hot electron distribution function. During the thermalization, the shape of the distribution function does not deviate much from that of the Maxwellian, while the mean energy shows only limited increase or even slight decrease depending on the strength of the self-limiting sink. Such behavior is due to the competition between the preferential depletion of lower energy hot electrons by collision and the selective depletion of higher energy hot electrons by the self-limiting sink.

Based on this observation, we modelled the finite-time thermalization in JOREK fluid simulations by two species, one representing the hot-tail electrons the density and temperature of which are functions of the normalized magnetic flux, the other is the cold bulk fluid electrons. The hot electron temperature is kept constant while their density decays according to the collisional and the self-limiting density sink, which are obtained by integrating over the hot electron distribution function. The ablation enhancement can then be estimated using the hot electron temperature and density. Incorporating such treatment, our JOREK axisymmetric simulations find that the hot-tail effect indeed cause stronger SPI assimilation and shallower penetration which is partly mitigated by the self-limiting effect. For ITER-like parameters, no significant effect is found on the assimilation, while the penetration is found to be somewhat shallower when the hot-tail effect is included, depending on the exact injection parameters. It is found that the injection velocity plays a role in determining the hot-tail effect strength, with slower injection showing weaker hot-tail influence since there is more time for thermalization to complete before the following fragments arrive. Reducing the longitudinal spreading of the fragment plume shows some self-regulated behavior, as the more compact fragments means less complete thermalization seen by fragments that follow the vanguard ones, while the self-limiting mitigation is also increased. Apart from the ITER-like parameters, an extreme case of Unshattered Pellet injection is also used to demonstrate the maximum ablation enhancement by the hot-tail electrons, which result in drastically larger assimilation and shallower penetration.

The conclusion that ITER-like SPIs do not exhibit strong hot-tail effect provides some reassurance on the validity of current mainstream fluid simulation of SPI in ITER. However, caution should be taken when considering extreme injection configurations such as unshattered pellets. Furthermore, a few effects that we did not take into account may affect the conclusion. For example, we have not taken into account the major radial plasmoid drift \cite{Rozhansky1995PPCF,Pegourie2007NF} which transports the ablated material outward along the major radius, reducing the density peak near the vanguard fragments. This would likely result in slower thermalization and thereby increasing the hot-tail effect. On the other hand, such drift will move the cold plasmoid away from the cooling front before they fully thermalize. Consequently, fluid simulations taking such drift effect into account could experience a warmer cooling front and thus stronger ablation. Hence, it is not trivial to estimate how much deviation between the hot-tail and the fluid description would the plasmoid drift cause. Another possibly important physics that is missing in this study is the 3D effect. First, in reality local unrelaxed density peak could develop around the fragment plume, resulting in enhanced thermalization. On the other hand, if the hot electron mean-free-path is longer than the size of such unrelaxed plasmoid cloud, the hot electrons will still ``see'' the flux-averaged density. In this case the thermalization should be the same as is studied in our axisymmetric case. Second, the flux tube volume $\gD V$ is greatly decreased on rational surfaces, which results in stronger self-limiting effect. Last, it could be envisioned that hot electrons from core could be released to the edge at the time of TQ or partial thermal collapses during the pre-TQ phase, causing additional ablation if their mean-free-path is longer than the connection length. The above effects are beyond the scope of our current study, and will be pursued in future works.

One future expansion of the present work is the inclusion of impurity species. The additional medium Z impurity such as neon would increase the collision rate thus provide faster thermalization. On the other hand, strong local impurity cooling may create local cold spots near the fragments in fluid simulation while in reality the mean free path of hot-tail electrons is longer than those cold spots. This will result in an artificially low ablation rate in the fluid simulations. Understanding how these two effects would balance each other will be left for future investigations.

Another issue we did not delve into is the electric field around the ablating fragments and its local alternation to the electron distribution function. Electric potential around the solid fragments and around the cold plasmoid may build up as the ambient electrons have a longer mean-free-path than the accompanying ions. The consequential electric field is not considered here in this study, and their impact on the electron distribution function and the ablation enhancement is left for future studies.

Nevertheless, the result of this study provides some insight to the thermalization dynamics of the hot electrons during SPI and provides a first estimation of the hot-tail impact to the important injection results such as assimilation and penetration, which determines the efficiency of disruption mitigation.

\vskip1em
\centerline{\bf Acknowledgments}
\vskip1em

  The authors thank B. N. Breizman, P. Aleynikov, E. Nardon and M. Lehnen for fruitful discussion. ITER is the Nuclear Facility INB no. 174. The views and opinions expressed herein do not necessarily reflect those of the ITER Organization. This publication is provided for scientific purposes only. Its contents should not be considered as commitments from the ITER Organization as a nuclear operator in the frame of the licensing process. Part of this work is supported by the National Natural Science Foundation of China under Grant No. 11905004, and the National MCF Energy R\&D Program of China under Grant No. 2019YFE03010001. This work is also partly supported by US Department of Energy grants DE-AC02-09CH11466. This work has been co-funded by the ITER Organization under the implementing agreement IO/IA/19/4300002053.  This work is partly carried out on Tianhe-3F operated by NSCC-TJ and partly used the Stellar cluster at Princeton University.

\end{document}